\documentclass[a4paper,12pt]{article}

\usepackage[english]{babel}
\usepackage{amsmath, amsthm, amsfonts, amssymb,graphicx}
\usepackage[mathscr]{eucal}

\theoremstyle{plain}
\newtheorem{theorem}{Theorem}[section]

\newtheorem{lemma}{Lemma}[section]

\theoremstyle{remark}
\newtheorem{remark}{Remark}[section]

\numberwithin{equation}{section}

\newcommand{\centsect}[1]
{

\bigskip

\addtocounter{section}{1}
\noindent{\large\textbf{\arabic{section}. #1}}
\setcounter{equation}{0}\setcounter{theorem}{0}
\setcounter{lemma}{0}\setcounter{remark}{0}\medskip }

\newcommand{\centsectn}[1]
{

\bigskip

\noindent{\large\textbf{#1}}\setcounter{equation}{0}
\setcounter{theorem}{0}\setcounter{lemma}{0}\setcounter{remark}{0}
\medskip}

\def\th{\theta}

\def\e{\varepsilon}
\def\g{\gamma}
\def\G{\Gamma}
\def\l{\lambda}
\def\p{\partial}
\def\D{\Delta}

\def\a{\alpha}

\def\t{\widetilde}

\def\d{\delta}
\def\L{\Lambda}
\def\z{\zeta}
\def\vs{\varsigma}

\def\vt{\vartheta}

\allowdisplaybreaks

\begin{document}

\begin{center}
{\Large{On a model boundary value problem for Laplacian with
frequently alternating type of boundary condition}}

\bigskip

{\large{Denis I. Borisov}}

\bigskip

The Bashkir State Pedagogical University, October Rev. St, 3a,
450000, Ufa, Russia

\medskip
tel. 7-3472-317827, fax 7-3472-229034

\medskip
E-mail: BorisovDI@ic.bashedu.ru, BorisovDI@bspu.ru
\end{center}

\bigskip

\begin{center}
\large{Abstract}
\end{center}

Model two-dimensional singular perturbed eigenvalue problem for
Laplacian with frequently alternating type of boundary condition
is considered. Complete two-parametrical asymptotics for the
eigenelements are constructed.

\newpage

\centsectn{Introduction}

Elliptic boundary value problems with frequently alternating type
of boundary condition are mathematical models used in various
applications. We briefly describe the formulation of these
problems. In a given bounded domain with a smooth or a piecewise
smooth boundary an elliptic equation is considered. On the
boundary one selects a subset depending on a small parameter and
consisting of a large number of disjoint parts. The measure of
each part tends to zero as the small parameter tends to zero,
while the number of these parts increases infinitely. On the
subset described the Dirichlet boundary condition is imposed,
whereas the Neumann boundary condition is imposed on the rest part
of the boundary. There is a number of papers devoted to averaging
of such problems (see, for instance, \ref{Fr}--\ref{OC}). The main
objective of these works was to describe limiting (homogenized)
problems. The case of periodic alternating of boundary conditions
was investigated in \ref{Ch}, \ref{LP}, while the nonperiodic one
was treated in \ref{Fr}, \ref{OC}. The main result of these works
can be formulated as follows. The form of limiting problem
(namely, type of boundary condition) depends of the relation
between measures of parts of the boundary with different types of
boundary conditions.

Further studying of the boundary value problems with frequently
alternating boundary conditions was carried out in two directions.
First direction consists in the estimates for degree of
convergence under minimal number of restrictions to the structure
of alternating of boundary conditions (\ref{Ch},
\ref{OC}--\ref{Dr}). Another direction in studying of these
problems is a constructing the asymptotics expansions of
solutions. Present paper develops exactly this direction.

In this paper we study a two-dimensional singular perturbed
eigenvalue problem for Laplace operator in a unit circle $D$
with center at the origin. On the boundary of the circle $D$ we
select a periodic subset $\g_\e$ consisting of $N$ disjoint
arcs, length of each arc equals $2\e\eta$, where $N\gg1$ is an
integer number, $\e=2N^{-1}$, $\eta=\eta(\e)$, $0<\eta<\pi/2$.
Each of these arcs can be obtained from an neighbouring one by
rotation about the origin through the angle $\e\pi$ (cf.
figure). On $\g_\e$ we  impose the Dirichlet boundary condition
and the Neumann boundary condition is considered on the rest
part of the boundary. From \ref{Fr}, \ref{Ch} it follows that
the main role in determination of limiting problem belongs to
the limit $\lim \limits_{\e\to0} (\e\ln\eta(\e))^{-1}=-A$. If
$A\ge0$, then the limiting problem is either the Robin problem
($A>0$) or the Neumann problem ($A=0$). The assumption $\lim
\limits_{\e\to0} (\e\ln\eta(\e))^{-1}=-A$ does not define the
function $\eta(\e)$ uniquely; clear, it is equivalent to the
equality $\eta(\e)=\exp\left(-\frac{1}{\e(A+\mu)}\right)$, where
$\mu=\mu(\e)$ is an arbitrary function tending to zero as
$\e\to0$, and also, $A+\mu>0$ for $\e>0$. Thus, the problem
studied contains actually two parameters, $\e$ and $\mu$. In
paper \ref{DU} complete power (on $\e$) asymptotics for the
eigenelements of the perturbed problem were constructed in the
case of the Neumann limiting problem ($A=0$) under an additional
assumption $\mu(\e)=A_0\e$, $A_0=\mathrm{const}>0$.

\begin{figure}[tb]
\begin{center}
\noindent
\includegraphics[width=7.4 cm,height=7.3 cm]{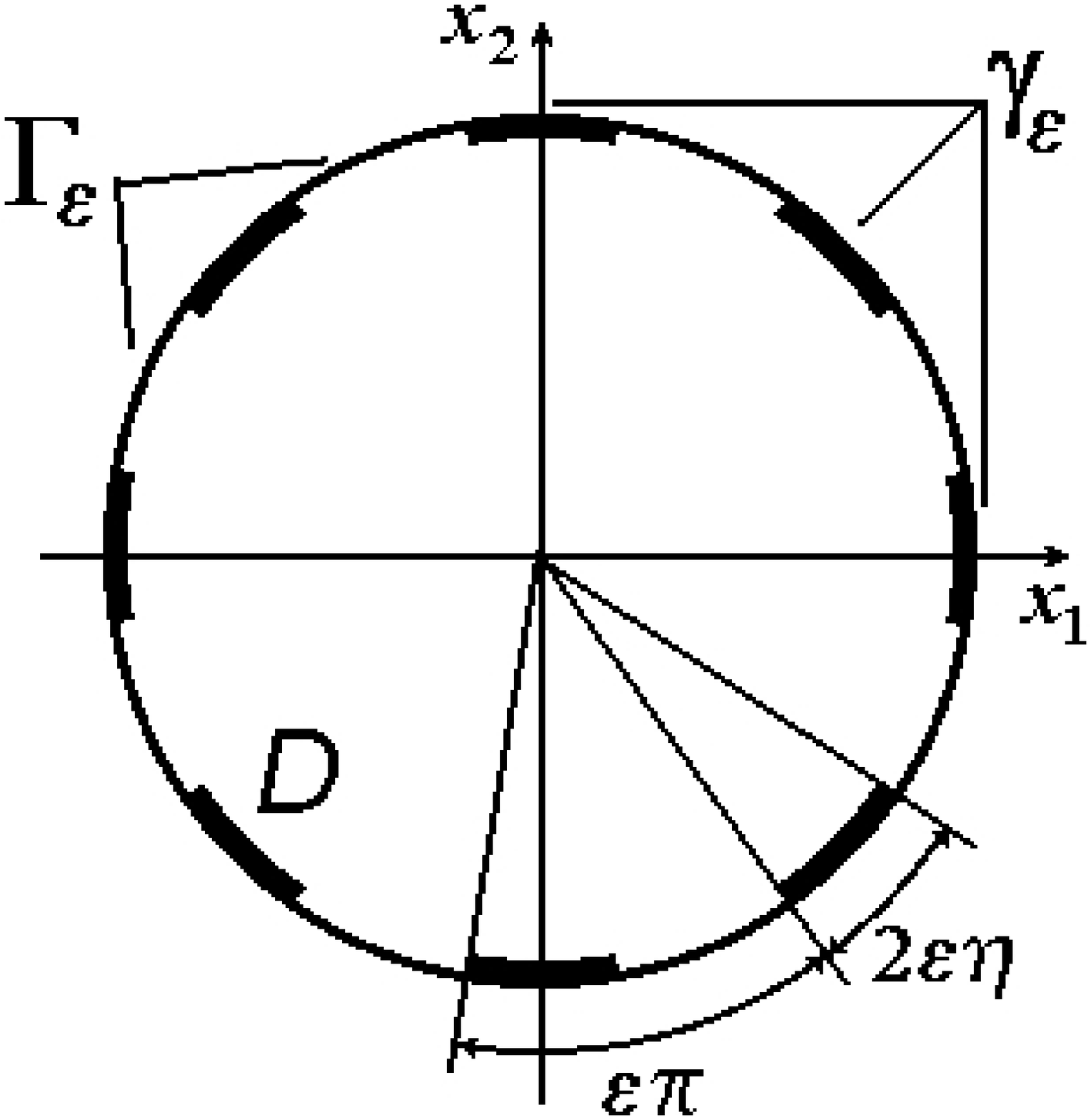}

\medskip

Figure.
\end{center}
\end{figure}

In this paper we study the case of limiting Neumann or Robin
problem ($A\ge0$) without any additional assumptions for
$\eta(\e)$. On the basis of the method of matched asymptotics
expansions \ref{ME}, the method of composite expansions \ref{CE}
and the multiscaled method \ref{MS} we obtain complete
two-parametrical (on $\e$ and $\mu$) asymptotics for the
eigenelements of the perturbed problem. Employing the
asymptotics expansions for the eigenvalues, we prove that the
perturbed problem has only simple and double eigenvalues, and we
show criterion distinguishing these cases.

\centsect{The problem and main results}

Let $x=(x_1,x_2)$ be the Cartesian coordinates, $(r,\th)$ be the
associated polar coordinates, $\G_\e=\p
D\backslash\overline{\g}_\e$. Without loss of generality we may
assume that the set $\g_\e$ is symmetric with respect to the
axis $Ox_1$. We study singular perturbed eigenvalue problem
\begin{align}
-{}&\D\psi_\e=\l_\e\psi_\e,\quad x\in D,\label{1.1}
\\
{}&\psi_\e=0,\quad x\in\g_\e,\qquad\frac{\p\psi_\e}{\p
r}=0,\quad x\in\G_\e. \label{1.2}
\end{align}
>From \ref{Fr}, \ref{Ch} it follows that in the case $A\ge0$ the
eigenelements of the perturbed problem converge to the
eigenelements of the following limiting problem
\begin{equation}
-\D\psi_0=\l_0\psi_0,\quad x\in D,\qquad\left(\frac{\p}{\p
r}+A\right)\psi_0=0, \quad x\in \p D.\label{1.3}
\end{equation}
The eigenfunctions converge strongly in $L_2(D)$ and weakly in
$H^1(D)$. Total multiplicity of the perturbed eigenvalues
converging to a $p$-multiply eigenvalue equals $p$.

It is well known fact that the eigenvalues of the problem
(\ref{1.3}) coincide with the roots of the equation
\begin{equation}
\sqrt{\l_0} J'_n\left(\sqrt{\l_0}\right)+A
J_n\left(\sqrt{\l_0}\right)=0,\label{1.4}
\end{equation}
where $J_n$ are Bessel functions of integer order  $n\ge0$, and
associated eigenfunctions are defined by the equalities
$\psi_0=J_0\left(\sqrt{\l_0}r\right)$ (for $n=0$) and
$\psi_0^\pm=J_n\left(\sqrt{\l_0}r\right)\phi^\pm(n\th)$ (for
$n>0$), $\phi^+=\cos$, $\phi^-=\sin$.
\begin{remark}\label{rm1.2}
It should be stressed that the problem (\ref{1.3}) can have
eigenvalues of various multiplicity, including multiplicity more
than two. This situation takes place because for some values of
$A$ there exists $\l_0$ being root of equation (\ref{1.4}) for
different $n$ simultaneously. The proof of existence of such $A$
is given in Appendix.
\end{remark}

This paper is devoted to the proof of the following statement.

\begin{theorem}\label{th1.1}
Let $\l_0$ be a root of the equation (\ref{1.4}) for $n\ge0$.
Then there exists an eigenvalue $\l_\e$ of the perturbed problem
converging to $\l_0$ and satisfying asymptotics
\begin{equation}
\l_\e=\L_0(\mu)+\sum\limits_{i=3}^{M-1}\e^i\L_i(\mu)+
O(\e^M(A+\mu)),\label{1.5}
\end{equation}
for any $M\ge3$, where $\L_0(\mu)$ is the root of the equation
\begin{align}
{}& \sqrt{\L_0} J'_n\left(\sqrt{\L_0}\right)+(A+\mu)
J_n\left(\sqrt{\L_0}\right)=0,\quad \L_0(0)=\l_0,\label{1.6}
\\
{}&
\begin{aligned}
{}&\L_3(\mu)=-\frac{\z(3)}{4}\frac{(A+\mu)^2\left(\L_0(\mu)+2n^2\right)
\L_0(\mu)}{\L_0(\mu)-n^2+(A+\mu)^2},
\\
{}&
\L_4(\mu)=\frac{\pi^4}{5760}\frac{(A+\mu)^2\left(8\L_0(\mu)+1\right)
\L_0(\mu)}{\L_0(\mu)-n^2+(A+\mu)^2},
\end{aligned}\label{1.8}
\end{align}
$\z(t)$ is the Riemann zeta function. The functions $\L_i(\mu)$,
$i\ge0$, are holomorphic on $\mu$;  for  $A=0$ and $i\ge3$ the
representations $\L_i(\mu)=\mu^2\widetilde{\L}_i(\mu)$ hold,
where $\widetilde{\L}_i(\mu)$ are holomorphic on $\mu$
functions. The eigenvalue $\l_\e$ is simple, if $n=0$, and it is
double, if  $n>0$. The asymptotics of the associated
eigenfunctions have the form (\ref{2.34}) for $n=0$ and
(\ref{3.1}) for $n>0$.
\end{theorem}
\begin{remark}\label{rm1.1}
It is known (\ref{KF}) that for $n\ge0$ the functions $J_n(t)$
and $J'_n(t)$ are positive at the points $t\in(0,n]$. For this
reason, the least root of the equation (\ref{1.4}) exceeds
$n^2$, what and (\ref{1.5}) imply the same for $\L_0(\mu)$,
i.e., the denominators in (\ref{1.8}) are nonzero. If
$A=n=\l_0=0$, then $\L_0>0$ and $\mu>0$, and the denominators in
(\ref{1.8}) are nonzero again.
\end{remark}

\begin{remark}\label{rm1.3}
It should be stressed that Theorem~\ref{th1.1} can be applied to
each eigenvalue of the perturbed problem. If $\l_0$ is a root of
the equation (\ref{1.4}) only for one value of $n$, then
Theorem~\ref{th1.1} implies immediately that only one perturbed
eigenvalue converges to $\l_0$ and this perturbed eigenvalue is
simple or double. if $\l_0$ is a root of the equation
(\ref{1.4}) for some values $n=n_i$, $i=1,\ldots,m $, $m\ge2$,
then for this case below it will be shown (see
Lemma~\ref{lm4.4}) that asymptotic series
(\ref{1.5})--(\ref{1.8}) do not coincide for different $n$, and
for this reason, exactly $m$ perturbed eigenvalues (that are
simple or double) converge to $\l_0$ that have asymptotics
(\ref{1.5})--(\ref{1.8}) with $n=n_i$, $i=1,\ldots,m $.
\end{remark}

This paper has the following structure. In two next sections we
formally construct asymptotics for the eigenvalues converging to
the roots of the equation (\ref{1.4}). Also we formally
construct the asymptotics for the associated eigenfunctions. We
separate the cases $n=0$ and $n>0$, the former is considered in
the second section, while the latter is studied in the third
one. However, the results of the second and third section do not
guarantee that the asymptotic series constructed formally are
really asymptotics of the eigenelements of the perturbed
problem. In the fourth section we carry out the justification of
the asymptotics, i.e., we prove that the asymptotic series
formally constructed do coincide with the asymptotics of the
eigenelements of the perturbed problem. As it has been already
mentioned in Remark~\ref{rm1.2}, in Appendix we prove  the
existence of positive $A$ for which there exists $\l_0$ being
root of the equation (\ref{1.4}) for different $n$
simultaneously.

\centsect{Formal construction of the asymptotics for the case
$n=0$}

In this section on the basis of the method of composite
expansions and the method of matched asymptotic expansions we
formally construct the asymptotics for an eigenvalue $\l_\e$,
converging to a root $\l_0$ of the equation (\ref{1.4}) with
$n=0$, and also, the asymptotics for the associated
eigenfunction $\psi_\e$.

At first, we briefly describe the scheme of construction. We
seek for the asymptotics of the eigenvalue as the series
(\ref{1.5}). It easily seen that the function
\begin{equation*}
\psi^{ex}_\e(x)=J_0\left(\sqrt{\l_\e}r\right),
\end{equation*}
is a solution of the equation (\ref{1.1}) for each $\l_\e$. At
the same time, it does not satisfy boundary condition
(\ref{1.2}). In order to satisfy homogeneous Neumann boundary
condition on $\G_\e$, using the method of composite expansions,
we construct a boundary layer in the vicinity of the boundary of
the circle $D$. This layer is constructed in the form of the
asymptotic series
\begin{equation}\label{2.2}
\psi^{mid}_\e(\xi)=\sum_{i=1}^\infty\e^i v_i(\xi,\mu),
\end{equation}
where $\xi=(\xi_1,\xi_2)=(\th\e^{-1},(1-r)\e^{-1})$ are
''scaled'' variables. However, the employment of only the method
of composite expansions does not allow to satisfy the
homogeneous Dirichlet boundary condition on $\G_\e$
simultaneously. In order to obtain the homogeneous Dirichlet
boundary condition, we apply the method of matched asymptotics
expansions in a neighbourhood of the points $x_m=(\cos\e\pi
m,\sin\e\pi m)$, $m=0,\ldots,N-1$, where we construct
asymptotics for the eigenfunction in the form
\begin{equation}\label{2.3}
\psi^{in}_\e\left(\vs^m\right)=\sum_{i=1}^\infty\e^i\left(
w_{i,0}\left(\vs^m,\mu\right)+\e\eta w_{i,1}
\left(\vs^m,\mu\right)\right),
\end{equation}
$\vs^m=\left(\vs^m_1,\vs^m_2\right)=\left((\xi_1-m\pi)\eta^{-1},
\xi_2\eta^{-1}\right)$. Note that the functions $w_{i,1}$ in
(\ref{2.3}) are not needed for formal construction of power (on
$\e$) asymptotics. They play an auxiliary role in the proof of
Theorem~\ref{th2.1}, which is used in justification of the
asymptotics in the fourth section.

The objective of this section is to determine the coefficients
of the series (\ref{1.5}), (\ref{2.2}) and (\ref{2.3}). We shall
obtain the explicit formulae for these quantities.

Let us proceed to construction. In accordance with the method of
composite expansions we postulate the sum of the functions
$\psi_\e^{ex}$ and $\psi_\e^{mid}$ to satisfy the homogeneous
boundary condition everywhere on the boundary $\p D$ except the
points $x_k$, i.e.,
\begin{equation*}
\sqrt{\l_\e} J'_0\left(\sqrt{\l_\e}\right)-
\frac{1}{\e}\frac{\p}{\p\xi_2}\psi_\e^{mid}=0,\quad \xi\in\G^0,
\end{equation*}
where $\G^0$ is the axis $O\xi_1$ without points  $(\pi k,0)$,
$k\in\mathbb{Z}$. Replacing $\l_\e$, $\psi_\e^{mid}$ by the series
(\ref{1.5}), (\ref{2.2}) in the equality obtained, expanding the
first term in Taylor series with respect to $\e$, and equaling to
zero the coefficients of powers of $\e$, we deduce boundary
conditions for the functions $v_i$:
\begin{align}
{}&\frac{\p v_i}{\p\xi_2}=\a_i, \quad \xi\in\G^0,\qquad
\a_i=\a_i(\L_0,\ldots,\L_{i-1}),\label{2.4}
\\
{}&\ \begin{aligned}
{}&\a_1=\sqrt{\L_0}J'_0\left(\sqrt{\L_0}\right),\qquad\a_2=\a_3=0,
\\
{}&\a_i=-\frac{1}{2}J_0\left(\sqrt{\L_0}\right)
\L_{i-1}+f_i,\quad i\geq4,\qquad f_4=f_5=0.
\end{aligned}\label{2.5}
\end{align}
Here $f_i=f_i(\L_0,\ldots,\L_{i-4})$ are polynomials on
variables $\L_1,\ldots,\L_{i-4}$ with holomorphic on $\L_0$
coefficients, moreover, $f_i(\L_0,0,\ldots,0)=0$. Let us deduce
the equations for the functions $v_i$. In order to do it, we
substitute $\psi_\e^{mid}$ and $\l_\e$ in the equation
(\ref{1.1}), and then pass to the polar coordinates what implies
the equation
\begin{equation*}
\left(r^2\frac{\p^2}{\p r^2}+r\frac{\p}{\p
r}+\frac{\p^2}{\p\th^2}+r^2\l_\e\right)\psi_\e^{mid}=0.
\end{equation*}
Replacing $\l_\e$ and $\psi_\e^{mid}$ by the series (\ref{1.5})
and (\ref{2.2}) in this equation, passing to the variables $\xi$
and equaling to zero coefficients of powers of $\e$, we can write
\begin{align}\label{2.6}
\D_\xi
v_i=&F_i\equiv\mathcal{L}_i(v_1,\ldots,v_{i-1}),\quad\xi_2>0,
\\
\mathcal{L}_i(v_1,\ldots,v_{i-1})=&\sum\limits_{k=0}^1
\sum\limits_{j=1}^2\mathsf{a}_{kj}\xi_2^{k+j-1}\frac{\p^j}{\p\xi_2^j}
v_{i-k}+\sum\limits_{k=0}^2 \mathsf{a_k}\xi_2^k
\sum\limits_{j=1}^{i-k-2}\L_{i-j-k-2}v_j,\nonumber
\end{align}
where
$\mathsf{a}_{11}=\mathsf{a}_{12}=\mathsf{a}_0=\mathsf{a}_2=-1$,
$\mathsf{a}_{01}=1$, $\mathsf{a}_1=\mathsf{a}_{02}=2$,
$v_{-1}=v_0=0$, $\L_1=\L_2=0$. The relations (\ref{2.4}),
(\ref{2.6}) are a recurrence system of boundary value problems
for the functions $v_i$. According to the method of composite
expansions, we are to seek its solutions exponentially decaying
as $\xi_2\to+\infty$. We shall obtain the explicit formulae for
$v_i$; for this we use the following auxiliary statements.

We indicate by $\mathcal{V}$ the space of $\pi$-periodic on
$\xi_1$ functions uniformly exponentially decaying as
$\xi_2\to+\infty$ together with all their derivatives, and
belonging to $C^\infty\left(\{\xi:\xi_2>0\}\cup\G^0\right)$. By
$\mathcal{V}^+$ ($\mathcal{V}^-$) we denote the subset of
$\mathcal{V}$ containing even (odd) on $\xi_1$ functions. We
introduce the operators $\mathcal{A}_k$, $k\ge0$ is an integer
number; their action on a function $u\in\mathcal{V}$ reads as
follows
\begin{equation*}
\mathcal{A}_0[u](\xi)=u(\xi),\quad\mathcal{A}_k[u](\xi)=
\int\limits^{+\infty}_{\xi_2}t \mathcal{A}_{k-1}[u](\xi_1,t)\,dt.
\end{equation*}
By definition of the spaces $\mathcal{V}$, $\mathcal{V}^+$ and
$\mathcal{V}^-$ and the definition of the operators
$\mathcal{A}_k$, one can check that
$\mathcal{A}_k:\mathcal{V}\to\mathcal{V}$ and
$\mathcal{A}_k:\mathcal{V}^\pm\to\mathcal{V}^\pm$.
\begin{lemma}\label{lm2.1}
For each $k\geq0$ the equalities
\begin{equation*}
\D_\xi\mathcal{A}_k[u]=-2k\mathcal{A}_{k-1}[u]+\mathcal{A}_k\left[\D_\xi
u\right]
\end{equation*}
hold.
\end{lemma}

\noindent\textbf{Proof.} Clear, for each function
$u\in\mathcal{V}$ we can write
\begin{equation*}
\frac{\p^{\mathsf{m_1}+\mathsf{m_2}}}
{\p\xi_1^\mathsf{m_1}\p\xi_2^\mathsf{m_2}}
\int\limits^{+\infty}_{\xi_2}u(\xi_1,t)\,dt
=\int\limits^{+\infty}_{\xi_2}
\frac{\p^{\mathsf{m_1}+\mathsf{m_2}}} {\p\xi_1^\mathsf{m_1}\p
t^\mathsf{m_2}} u(\xi_1,t)\,dt,\quad \xi_2>0,
\end{equation*}
where $\mathsf{m_1},\mathsf{m_2}\in\mathbb{Z_+}$, what yields
\begin{equation*}
\D_\xi\mathcal{A}_0[u]=\D_\xi u,\quad \D_\xi\mathcal{A}_k[u]=
\int\limits^{+\infty}_{\xi_2}t\D\mathcal{A}_{k-1}
[u](\xi_1,t)\,dt-2\mathcal{A}_{k-1}[u].
\end{equation*}
Employing the equalities obtained by induction, it is easy to
prove the lemma.

We set $\Pi=\{\xi: -\pi/2<\xi_1<\pi/2, \xi_2>0\}$, $(\rho,\vt)$
are the polar coordinates associated with the variables $\xi$.

\begin{lemma}\label{lm2.2} Let the function $F(\xi)\in\mathcal{V}^+$
has infinitely differentiable asymptotics
\begin{equation*}
F(\xi)=\a\rho^{-1}\sin 3\vt+O(\ln\rho),\quad\rho\to0,
\end{equation*}
and there exists a natural number $k$, such that $\D_\xi^k
F\equiv0$ for $\xi_2>0$. Then the function
\begin{equation}
v=-\sum\limits_{j=1}^k\frac{1}{2^j
j!}\mathcal{A}_j\left[\D_\xi^{j-1}F\right]\label{2.7}
\end{equation}
is a solution of the boundary value problem
\begin{equation}
\D_\xi v=F,\quad\xi_2>0,\qquad\frac{\p
v}{\p\xi_2}=0,\quad\xi\in\G^0,\label{2.8}
\end{equation}
belonging to $H^1(\Pi)\cap\mathcal{V}^+$, and having infinitely
differentiable asymptotics
\begin{equation}
v(\xi)=v(0)+\frac{1}{2}\a\xi_2^3\rho^{-2}+O(\rho^2\ln\rho),
\quad\rho\to0.\label{2.9}
\end{equation}
\end{lemma}
\noindent\textbf{Proof.} Since $F\in\mathcal{V}^+$, then,
obviously, $\D^j_\xi F\in\mathcal{V}^+$, and, therefore, each
term in the right hand side of (\ref{2.7}) belongs to
$\mathcal{V}^+$, what implies $v\in\mathcal{V}^+$. Let us check
that the function $v$ defined by the equality (\ref{2.7}) is
really a solution of the boundary value problem (\ref{2.8}).
Indeed, for each point $\xi\in\G^0$ we have
\begin{equation*}
\frac{\p
v}{\p\xi_2}\Big|_{\xi\in\G^0}=\sum\limits_{j=1}^k\frac{1}{2^j
j!}\left(\xi_2\mathcal{A}_{j-1}\left[\D_\xi^{j-1}F\right]\right)
\Big|_{\xi\in\G^0}=0.
\end{equation*}
For $\xi\in\Pi$, applying the Laplace operator to $v$, using
Lemma~\ref{lm2.1}, and employing the equality  $\D^k_\xi
F\equiv0$, we get
\begin{equation*}
\D_\xi v=-\sum\limits_{j=1}^k\frac{1}{2^j
j!}\mathcal{A}_j\left[\D_\xi^j F\right]+\sum\limits_{j=1}^k
\frac{2j}{2^j j!}\mathcal{A}_{j-1}\left[\D_\xi^{j-1}F\right]=F.
\end{equation*}
We proceed to the proof of the asymptotics (\ref{2.9}). Let a
function $U(\xi)\in\mathcal{V}^+$ have differentiable
asymptotics
\begin{equation}
U(\xi)=O\left(\rho^{-p}\ln^q\rho\right),\quad\rho\to0,\quad
p,q\in\mathbb{Z},\quad p,q\ge0.\label{2.10}
\end{equation}
We set $u(\xi)=\mathcal{A}_1[U](\xi)$. As $U\in\mathcal{V}^+$,
then the representation
\begin{equation}
u(\xi)=\int\limits^a_{\xi_2} tU(\xi_1,t)\,dt+u_1(\xi),\label{2.11}
\end{equation}
is true, where $u_1\in\mathcal V^+\cap C^\infty
(\{\xi:\xi_2\geq0\})$, $a$ is a fixed sufficiently small number.
It is obvious that
\begin{equation}\label{2.12}
u_1(\xi)=u_1(0)+O(\rho^2),\quad\rho\to0.
\end{equation}
Now we replace the function $U$ by its asymptotics (\ref{2.10})
in (\ref{2.11}). After that the integral in (\ref{2.11}) can be
calculated explicitly, from what and (\ref{2.12}) it follows
that
\begin{equation}
u(\xi)=O\left(\ln^{q+1}\rho\right),\quad p=2,\qquad
u(\xi)=O\left(\rho^{-p+2}\ln^q\rho\right),\quad p>2,\label{2.13}
\end{equation}
as $\rho\to0$. For $p=0,1$ one can see that
\begin{equation}
\begin{aligned}
u(\xi)-&u(0)=\int\limits_{\xi_2}^a t
\left(U(\xi_1,t)-U(0,t)\right)\,dt-\int\limits_0^{\xi_2}
tU(0,t)\,dt+O\left(\rho^2\right)=\\
&=\int\limits_0^{\xi_1}\int\limits_{\xi_2}^a t_2\frac{\p}{\p
t_1} U(t_1,t_2)\,dt_2
dt_1+O\left(\rho^{-p+2}\ln^q\rho\right)=O\left(\rho^{-p+2}\ln^q\rho\right),
\end{aligned}\label{2.14}
\end{equation}
as $\rho\to0$.  For the function $F$ we have the equalities as
$\rho\to0$
\begin{equation*}
\D_\xi F=-8\a\rho^{-3}\sin3\vt+O(\rho^{-2}\ln\rho),\qquad\D_\xi^j
F=O(\rho^{-2j}\ln\rho), \quad j\ge2,
\end{equation*}
which and (\ref{2.7}), (\ref{2.10}), (\ref{2.13}), (\ref{2.14})
and definition of the operators $\mathcal{A}_j$ imply the
asymptotics (\ref{2.9}). In view of latter and the inclusion
$v\in\mathcal{V}$ we conclude that $v\in H^1(\Pi)$. The proof is
complete.

Let $X(\xi)=\mathrm{Re}\,\ln\sin z+\ln 2-\xi_2$, where
$z=\xi_1+\mathrm{i}\xi_2$  is a complex variable. By direct
calculations we check that $X\in\mathcal{V}^+$ is a harmonic
function as $\xi_2>0$, satisfying the boundary condition
\begin{equation*}
\frac{\p X}{\p\xi_2}=-1,\quad\xi\in\G^0,
\end{equation*}
and having differentiable asymptotics
\begin{equation}\label{2.16}
X(\xi)=\ln\rho+\ln 2-\xi_2+O\left(\rho^2\right),\quad\rho\to0.
\end{equation}

The lemmas proved enable us to solve the system of the problems
(\ref{2.4}), (\ref{2.6}).

\begin{lemma}\label{lm2.3}
For each sequence $\left\{\L_i(\mu)\right\}_{i=0}^\infty$,
$\L_1(\mu)=\L_2(\mu)=0$ there exist solutions of the boundary
value problems (\ref{2.4}), (\ref{2.6}) defined by formula
(\ref{2.7}) with $F=F_i$, $k=k_i$, where $k_i$ are some natural
numbers. For the functions $v_i$ the representations
\begin{equation}\label{2.17}
\begin{aligned}
{}& v_i(\xi,\mu)=\t v_i(\xi,\mu)-\a_i X(\xi),
\\
{}&\t v_i(\xi,\mu)=\sum\limits_{j=1}^{M_i}
a_{ij}\left(\a_1,\L_0,\ldots, \L_{i-2}\right) v_{ij}(\xi),
\end{aligned}
\end{equation}
hold, where $v_{ij}\in H^1(\Pi)\cap\mathcal{V}^+$, $a_{ij}$ are
polynomials on $\L_1,\ldots,\L_{i-2}$ with holomorphic on $\L_0$
and $\a_1$ coefficients,
$a_{ij}\left(0,\L_0,0,\ldots,0\right)=0$. The equalities
$M_1=0$, $M_2=1$, $M_3=2$, $M_4=3$,
\begin{equation}\label{2.18}
\begin{aligned}
{}& a_{21}=a_{31}=a_{41}=-\a_1,\quad a_{32}=-\a_1\L_0,\quad
a_{42}=\frac{\a_1\left(6\L_0+1\right)}{24},
\\
{}& a_{43}=\frac{\a_1\left(8\L_0+1\right)}{32},\quad
v_{21}=\frac{1}{2}\xi_2^2\frac{\p X}{\p\xi_2},\quad
v_{31}=\frac{1}{8}\xi_2^4\frac{\p^4
X}{\p\xi_2^2}+\frac{1}{6}\xi_2^3\frac{\p X}{\p\xi_2},\\ &
v_{32}=\frac{1}{2}\mathcal{A}_1[X],\quad
v_{41}=\frac{1}{48}\xi_2^6\frac{\p^3 X}{\p\xi_2}+
\frac{2}{15}\xi_2^5\frac{\p^2 X}{\p\xi_2^2}+
\frac{1}{16}\xi_2^4\frac{\p X}{\p\xi_2},\\ &v_{42}=\xi_2^3
X,\quad v_{43}=\mathcal{A}_1[\xi_2
X]+\xi_2^2\int\limits_{\xi_2}^{+\infty}X(\xi_1,t)\,dt.
\end{aligned}
\end{equation}
take place.  The asymptotics
\begin{equation}\label{2.19}
v_i(\xi,\mu)=-\a_i\left(\ln\rho+\ln 2-\xi_2\right)+\t
v_i(0,\mu)-\frac{1}{2}\a_{i-1}\xi_2^3\rho^{-2}+
O\left(\rho^2\ln\rho\right),
\end{equation}
are correct as $\rho\to0$, where $\a_0=0$.
\end{lemma}

\noindent\textbf{Proof.} The statement of the lemma for
$i=1,\ldots,4$ and the equalities (\ref{2.18}) are checked by
direct calculations. For $i\ge5$ we carry out the proof by
induction. Let the lemma is valid for $i<K$. Then, due to
(\ref{2.6}) and induction assumption we have the relation
\begin{equation*}
F_K=\sum\limits_{j=1}^K a_{Kj} F_{Kj},
\end{equation*}
where $F_{Kj}$ satisfy to all assumptions of Lemma~\ref{lm2.2},
and the functions
$a_{Kj}=a_{Kj}\left(\a_1,\L_0,\ldots,\L_{K-2}\right)$ posses all
the properties described in the statement of the lemma being
proved. Let $v_{Kj}$ be the solutions of the problem (\ref{2.8})
for $F=F_{Kj}$ defined in accordance with (\ref{2.7}). Then $\t
v_K\in H^1(\Pi)\cap\mathcal{V}^+$ is a solution of the equation
(\ref{2.6}) for $i=K$, satisfying the homogeneous Neumann
boundary condition  on $\G^0$. From this fact it follows that
the function $v_K$ defined in accordance with (\ref{2.17}) is
really a solution  of the boundary value problem (\ref{2.4}),
(\ref{2.6}) for $i=K$. Clear, the function $F_K$ satisfies the
hypothesis of Lemma~\ref{lm2.2} and has the asymptotics
\begin{equation*}
F_K=-\a_{K-1}\rho^{-1}\sin 3\vt+O(\ln\rho),\quad\rho\to0,
\end{equation*}
which and (\ref{2.9}) imply
\begin{equation*}
\t v_i(\xi,\mu)=\t v_i(0,\mu)-\frac{1}{2}\a_{i-1}\xi_2^3\rho^{-2}+
O\left(\rho^2\ln\rho\right), \quad\rho\to0.
\end{equation*}
Combining the last equality with (\ref{2.16}), (\ref{2.17}), we
obtain the asymptotics (\ref{2.19}). The proof is complete.

As it follows from the definition of the functions $v_j$, the
sum of $\psi_\e^{ex}$ and $\psi_\e^{mid}$ does not satisfy
homogeneous Dirichlet boundary condition on $\g_\e$. Moreover,
the functions $v_j$ have logarithmic singularities at the points
$x_k$. For this reason, we use the method of matched asymptotics
expansions for the construction of the asymptotics for the
eigenfunction in a neighbourhood of these points. We construct
this asymptotics in the form of the series (\ref{2.3}). The
functions $v_j$ being periodic on $\xi_1$, it is sufficient to
carry out the matching in the vicinity of the point $x_0=(1,0)$
and then to extend the results obtained for other points $x_k$.

We introduce the notation $\vs=\vs^0=\xi\eta^{-1}$. Let us
substitute the series (\ref{1.5}) and (\ref{2.3}) in
(\ref{1.1}), (\ref{1.2}), and calculate after that the
coefficients of the same powers of $\e$. As a result, we have
the following problems for $w_{i,j}$:
\begin{gather}
 \D_\vs w_{i,0}=0,\quad\vs_2>0,\qquad
w_{i,0}=0,\quad\vs\in\g^1,\quad
\frac{\p}{\p\vs_2}w_{i,0}=0,\quad\vs\in\G^1,\label{2.20}
\\
\begin{aligned}
{}& \D_\vs w_{i,1}
=\left(\frac{\p}{\p\vs_2}+2\vs_2\frac{\p^2}{\p\vs_2^2}
\right)w_{i,0},\quad\vs_2>0,
\\
{}& w_{i,1}=0,\quad\vs\in\g^1,\quad\frac{\p}{\p\vs_2}w_{i,1}=0,
\quad\vs\in\G^1,
\end{aligned}\label{2.21}
\end{gather}
where $\g^1$ is the interval $(-1,1)$ in the axis $\vs_2=0$, and
$\G^1$ is the complement of $\overline{\g}^1$ on the axis
$O\vs_1$. Next following the method of matched asymptotics
expansions, we calculate the asymptotics as $|\vs|\to\infty$ for
the functions $w_{i,j}$. We denote
\begin{align*}
{}& \l_{\e,K}=\L_0(\mu)+\sum\limits_{i=3}^{K}\e^i\L_i(\mu),\quad
\psi_{\e,K}^{ex}(x)=J_0\left(\sqrt{\l_{\e,K}}r\right),
\\
{}& \psi_{\e,K}^{mid}(\xi)=\sum\limits_{i=1}^{K+1}\e^i
v_i(\xi,\mu),\quad
\Psi_{\e,K}(x)=\psi_{\e,K}^{ex}(x)+\chi(1-r)\psi_{\e,K}^{mid}(\xi),
\end{align*}
where $\chi(t)$ is an infinitely differentiable cut-off function
equal to one as $t<1/3$ and to zero as $t>1/2$. Expanding in
Taylor series, we can write
\begin{align}
{}&
\begin{aligned}
J_0\left(\sqrt{\l_{\e,K}}r\right)=&\sum\limits_{i=0}^{K}\e^i
G_i(\L_0,\ldots,\L_i)+\e^{K+1}G^{(K)}_\e\left(\L_0,\ldots,
\L_K\right)-
\\
{}&-\e\xi_2\sqrt{\l_{\e,K}}J'_0\left(\sqrt{\l_{\e,K}}\right)
+O\left(\e^2\xi^2_2\right),
\end{aligned}\label{2.22}
\\
{}&\qquad
\begin{aligned}
{}&G_0=J_0\left(\sqrt{\L_0}\right),\qquad G_1=G_2=0,
\\
{}&G_i=\frac{ J'_0\left(\sqrt{\L_0}\right)}{2\sqrt{\L_0}}\L_i+
g_i,\quad i\ge3,\qquad g_3=g_4=0,
\end{aligned}\label{2.23}
\end{align}
where the functions $g_i=g_i\left(\L_0,\ldots,\L_{i-3}\right)$
are polynomials with respect to $\L_1$, \ldots, $\L_{i-3}$ with
holomorphic on $\L_0$ coefficients,
$g_i\left(\L_0,0,\ldots,0\right)=0$, $G^{(K)}_\e$ is a bounded
holomorphic on $\L_1,\ldots,\L_K$ functions,
$G^{(K)}_\e\left(\L_0,0,\ldots,0\right)=0$. From (\ref{2.19})
and the equality $\ln\eta=-\frac{1}{\e(A+\mu)}$ it follows that
\begin{equation}\label{2.24}
\begin{aligned}
v_i(\xi,\mu)=&\frac{1}{\e}\frac{\a_i}{A+\mu}-
\a_i\left(\ln|\vs|+\ln 2\right)+\a_i\xi_2+\t v_i(0,\mu)-
\\
{}&-\frac{1}{2}\eta\a_{i-1}\vs_2^3|\vs|^{-2}+
O\left(\eta^2|\vs|^2\ln|\vs|\right)
\end{aligned}
\end{equation}
as $\eta^{1/2}<\rho<2\eta^{1/2}$ (i.e., as
$\eta^{-1/2}<|\vs|<2\eta^{-1/2}$). We substitute (\ref{2.22})
and (\ref{2.24}) in the formula for $\Psi_{\e,K}$. Then, as
$\eta^{-1/2}<|\vs|<2\eta^{-1/2}$,
\begin{align}
{}&
\begin{aligned}
\Psi_{\e,M}(x)=&\sum\limits_{i=0}^{K}\e^i W_{i,0}(\vs,\mu)+\e
\eta\sum\limits_{i=1}^K\e^i W_{i,1}(\vs,\mu)+
\\
{}&+W^{(K)}_\e(x)+ O\left(\e\eta^2|\vs|^2\ln|\vs|\right),
\end{aligned}\nonumber
\\
{}& W_{i,0}(\vs,\mu)=-\a_i\left(\ln|\vs|+\ln 2\right)
+\frac{\a_{i+1}}{A+\mu}+\t v_i(0,\mu)+G_i,\quad\t
v_0=0,\label{2.26}
\\
{}&
W_{i,1}(\vs,\mu)=-\frac{1}{2}\a_i\vs_2^3|\vs|^{-2},\label{2.27}
\\
{}& W^{(K)}_\e(x)=\e^{K+1}\left(-\a_{K+1}\left(\ln|\vs|+\ln
2-\xi_2\right)+G^{(K)}_\e+\t v_{K+1}(0,\mu)\right)+
\e\mathsf{b}_K(\e)\xi_2,\nonumber
\\
{}&\quad\;\mathsf{b}_K(\e)=\sum\limits_{i=1}^K\e^{i-1}\a_i-
\sqrt{\l_{\e,K}}J'_0\left(\sqrt{\l_{\e,K}}\right).\nonumber
\end{align}
Observe, in view of definition of the functions $\a_i$, the
quantity $\mathsf{b}_K(\e)$ is small, namely,
$\mathsf{b}_K(\e)=O\left(\e^K\right)$. In accordance with the
method of matched asymptotics expansions, we must find the
solutions of (\ref{2.20}), (\ref{2.21}), satisfying the
asymptotics
\begin{equation}\label{2.28}
w_{i,j}(\vs,\mu)=W_{i,j}(\vs,\mu)+o\left(|\vs|^j\right),
\quad|\vs|\to\infty.
\end{equation}
We introduce the function
$Y(\vs)=\mathrm{Re}\,\ln\left(y+\sqrt{y^2-1}\right)$, where
$y=\vs_1+\mathrm{i}\vs_2$ is a complex variable. By definition,
$Y(\vs)$ is a solution of the problem (\ref{2.20}) and has the
asymptotics
\begin{equation}\label{2.29}
Y(\vs)=\ln|\vs|+\ln2+O\left(|\vs|^{-2}\right),\qquad
|\vs|\to\infty.
\end{equation}
>From the properties of the function $Y$, the asymptotics
(\ref{2.26}), (\ref{2.28}), (\ref{2.29}) and the problem
(\ref{2.20}) we deduce that
\begin{equation}\label{2.30}
w_{i,0}=-\a_i Y.
\end{equation}
Comparing the asymptotics for the function $w_{i,0}$ implied by
(\ref{2.29}), (\ref{2.30}) with the equalities (\ref{2.26}),
(\ref{2.28}), we conclude that
\begin{equation}\label{2.31}
\t v_i(0,\mu)+\frac{\a_{i+1}}{A+\mu}+G_i=0.
\end{equation}
>From the equality obtained for $i=0$ and from (\ref{2.5}),
(\ref{2.23}) it follows the equation (\ref{1.6}) for $\L_0$. The
condition $\L_0(0)=\l_0$ is obvious due to $\l_\e\to\l_0$. If
$\l_0\not=0$, then the holomorphy of $\L_0$ on $\mu$ is the
corollary to the implicit function theorem. If $\l_0=0$, then
$A=0$, and in this case the equation (\ref{1.6}) has a solution
of the form $\L_0(\mu)=\mu\widetilde{\L}_0(\mu)$, where
$\widetilde{\L}_0$ is a holomorphic on $\mu$ function,
$\widetilde{\L}_0(0)=2$. By Lemma~\ref{lm2.3} (see (\ref{2.17}),
(\ref{2.18})) we have: $\t v_1(0,\mu)=\t v_2(0,\mu)\equiv0$.
Employing these relations and the equalities
$\a_2=\a_3=G_1=G_2=0$ (see (\ref{2.5}), (\ref{2.23})) one can
check that the equality (\ref{2.31}) holds for $i=1,2$. Let us
consider the case $i\ge3$. Substituting the formulae
(\ref{2.5}), (\ref{2.31}) for $\a_{i+1}$ and $G_i$ into
(\ref{2.23}), then expressing $\L_i$ from what obtained, we
write
\begin{align}\label{2.32}
{}&\L_i(\mu)=\frac{2\L_0(\mu)\left(\t f_{i+1}(\mu)+(A+\mu)
\left(\t g_i(\mu)+\t
v_i(0,\mu)\right)\right)}{J_0(\sqrt{\L_0(\mu)})
\left(\L_0(\mu)+(A+\mu)^2\right)},
\\
{}& \t f_{i+1}(\mu)=
f_{i+1}\left(\L_0(\mu),\ldots,\L_{i-3}(\mu)\right), \quad \t
g_i(\mu)=g_i\left(\L_0(\mu),\ldots,\L_{i-3}(\mu)\right).\nonumber
\end{align}
Deducing this equality, in view of the equation (\ref{1.6}) we
replaced $J'_0\left(\sqrt{\L_0}\right)$ by the function
$-(A+\mu)J_0\left(\sqrt{\L_0}\right)/\sqrt{\L_0}$). Making
$i=3,4$ in (\ref{2.32}) and employing  (\ref{2.5}),
(\ref{2.17}), (\ref{2.18}), (\ref{2.23}) and the equalities
(\ref{GR})
\begin{equation}
\mathcal{A}_1[X](0)=-\frac{1}{4}\z(3),\quad \mathcal{A}_1[\xi_2
X](0)=-\frac{\pi^4}{360},\label{2.37}
\end{equation}
we get (\ref{1.8}) for $n=0$.

Let us prove that $\L_i$ are holomorphic on $\mu$. Since
$J_0\left(\sqrt\l_0\right)\not=0$, then the function
$J_0\left(\sqrt{\L_0(\mu)}\right)$ is holomorphic on  $\mu$ and
does not vanish for small $\mu\ge0$. If $\l_0\not=0$, then the
function $\left(\L_0(\mu)+(A+\mu)^2\right)$ also does not vanish
for $\mu\ge0$. In the case $\l_0=0$ (here $A=0$) the functions
$\L_0(\mu)$ and $\left(\L_0(\mu)+\mu^2\right)$ have a zero of
first order at the point $\mu=0$, so, for all possible values of
$\l_0$ and $A$ the quotient
\begin{equation*}
\displaystyle\frac{2\L_0(\mu)}{J_0\left(\sqrt{\L_0(\mu)}\right)
\left(\L_0(\mu)+(A+\mu)^2\right)}
\end{equation*}
is a holomorphic function as $\mu\ge0$. In view of the statement
of Lemma~\ref{lm2.3} for the functions $a_{ij}$ and of the
formula (\ref{2.17}) for the function $\t v_i$, the function $\t
v_i(0,\mu)$ is holomorphic on $\mu$, provided
$\L_0,\ldots,\L_{i-1}$ are holomorphic on $\mu$. Using this fact
and that the functions $f_{i+1}$ and $g_i$ are holomorphic on
$\L_0,\ldots,\L_{i-3}$, one can easy prove by induction that
$\L_i$ are holomorphic on $\mu$.

We proceed to the case $A=0$. For $i=3,4$ from  (\ref{1.8}) it
follows that $\L_i(\mu)=\mu^2\t\L_i(\mu)$, where $\t\L_i(\mu)$
are holomorphic on $\mu$ functions. Let us show the same for
$i\ge5$. Suppose that it is true for $i<M$. Since the functions
$f_{M+1}$, $g_M$, $\t v_M(0,\mu)$ are holomorphic on
$\L_0,\ldots,L_{M-1}$, $\a_1$ is holomorphic on $\L_0$,
$f_{M+1}\left(\L_0,0,\ldots,0\right)=g_M\left(\L_0,0,\ldots,0\right)=
a_{Mj}\left(0,\L_0,0,\ldots,0\right)=0$, then $\t
f_{M+1}(\mu)=\mu^2\t f^{(M+1)}(\mu)$, $\t g_M(\mu)=\mu^2\t
g^{(M)}(\mu)$, $\t v_M(0,\mu)=\mu\t v^{(M)}(\mu)$, where $\t
f^{(M+1)}(\mu)$, $\t g^{(M)}(\mu)$, $\t v^{(M)}(\mu)$ are
holomorphic on $\mu$ functions. By this fact and (\ref{2.32}) we
arrive at the desired representations.

Let us determine the functions $w_{i,1}$. By direct calculations
we check that the solutions of the problems (\ref{2.21}),
satisfying asymptotics (\ref{2.27}), (\ref{2.28}), have the form
\begin{equation}\label{2.33}
w_{i,1}=\frac{1}{2}\vs_2^2\frac{\p}{\p\vs_2}w_{i,0}.
\end{equation}

Thus, the formally constructed asymptotics for the eigenfunction
looks as follows
\begin{align}
{}&\begin{aligned}
\psi_\e(x)=&\left(\psi^{ex}_\e(x)+\chi(1-r)\psi^{mid}_\e(\xi)\right)
\chi_\e(x)+\\
{}&+\sum\limits_{m=0}^{N-1}\chi\left(\left|\vs^m\right|\eta^{1/2}
\right)\psi^{in}_\e \left(\vs^m\right),
\end{aligned}\label{2.34}
\\
{}&\chi_\e(x)=1-\sum_{m=0}^{N-1}\chi\left(\left|\vs^m\right|\eta^{1/2}\right).
\nonumber
\end{align}
We introduce the notations
\begin{align*}
\psi_{\e,K}(x)=&\Psi_{\e,K}(x)\chi_\e(x)+
\sum\limits_{m=0}^{N-1}\chi\left(\left|\vs^m\right|\eta^{1/2}
\right)\psi^{in}_{\e,K}\left(\vs^m\right),
\\
\psi^{in}_{\e,K}\left(\vs\right)=&\sum_{i=1}^K\e^i\left(
w_{i,0}(\vs,\mu) +\e\eta w_{i,1}(\vs,\mu)\right),
\\
\widetilde{\psi}_{\e,K}(x)=&\psi_{\e,K}(x)-R_{\e,K}(x),
\\
R_{\e,K}(x)=&\chi(1-r)
\left(\mathsf{b}_k(\e)(1-r)-\e^{K+1}\a_{K+1}X(\xi)\right)+
\\
{}&+\e^{K+1}G^{(K)}_\e+\e^{K+1}\t
v_{K+1}(0,\mu)-\e^K\frac{\a_{K+1}}{A+\mu}.
\end{align*}
We set $\|\bullet\|=\|\bullet\|_{L_2(D)}$.

\begin{theorem}\label{th2.1}
The functions $\psi_{\e,K}, \t\psi_{\e,K}\in H^1(D)\cap
C^\infty(D)$ converges to $\psi_0$ in $L_2(D)$ as $\e\to0$,
$\l_{\e,K}$ converges to $\l_0$,
$\left\|R_{\e,K}\right\|=O(\e^{K}(A+\mu))$. The functions
$\t\psi_{\e,K}$ and $\l_{\e,k}$ are the solutions of the problem
\begin{equation}\label{2.35}
-\D u_\e=\l u_\e+f,\quad x\in D,\quad u_\e=0,\quad
x\in\g_\e,\qquad \frac{\p u_\e}{\p r}=0,\quad x\in\G_\e,
\end{equation}
with $u_\e=\t\psi_{\e,K}$, $\l=\l_{\e,K}$, $f=f_{\e,K}$, where
$\left\|f_{\e,K}\right\|=O\left(\e^K(A+\mu)\right)$.
\end{theorem}
\begin{remark}\label{rm2.1}
The expressions of the form $O\left(\e^p(A+\mu)\right)$ in the
statement of this theorem should be interpreted in the following
way. For $A>0$ it means $O(\e^p)$, for $A=0$ it does
$O(\e^p\mu)$.
\end{remark}

\noindent\textbf{Proof.} The desired smoothness of $\psi_{\e,K}$
and $\t\psi_{\e,K}$ follows directly from the definition of
these functions and the smoothness of the functions
$\psi_{\e,K}^{ex}$, $v_i$ and $w_{i,j}$. It is obvious that
$\l_{\e,K}\to\l_0$, $\psi_{\e,K}, \t\psi_{\e,K}\to\psi_0$ as
$\e\to0$. By definition and properties of the quantities $\a_i$
we deduce that $\mathsf{b}_k(\e)=O\left(\e^{K}(A+\mu)\right)$,
from what and the definition of the functions $G^{(K)}_\e$ and
$\a_{K+1}$ and the smoothness of the function $\chi$ it follows
that $\left\|R_{\e,K}\right\|=O\left(\e^{K+1}(A+\mu)\right)$.

Since the function $\chi_\e(x)$ equals zero in a small
neighbourhood of the set $\g_\e$, and the functions $w_{ij}$
vanish on $\g^1$, then the function $\t\psi_{\e,K}$ satisfies
Dirichlet homogeneous boundary condition on $\g_\e$. By direct
calculations we check that for $x\in\G_\e$
\begin{align*}
\frac{\p}{\p r}\t\psi_{\e,K}(x)=&\chi_\e(x)\frac{\p}{\p
r}\Psi_{\e,K}(x)-\frac{\p}{\p r}R_{\e,K}(x)=\chi_\e(x)\Bigg(
\sqrt{\l_{\e,K}}J'_0\left(\sqrt{\l_{\e,K}}\right)-
\\
{}&\left. -\sum\limits_{i=1}^{K}\e^{i-1}\frac{\p
v_i}{\p\xi_2}\Big|_{\xi\in\G^0}+\e^K\a_{K+1}+\mathsf{b}_K(\e)
\right)=0.
\end{align*}

Applying the operator $-\left(\D+\l_{\e,K}\right)$ to the
function $\t\psi_{\e,K}(x)$, we obtain that
\begin{align*}
f_{\e,K}&=-\sum\limits_{i=1}^5 f^{(i)}_{\e,K},\qquad\text{where}
\\
f^{(1)}_{\e,K}(x)&=-\chi_\e(x)\left(\D+\l_{\e,K}\right)
R_{\e,K}(x),
\\
f^{(2)}_{\e,K}(x)&=\chi(1-r)\chi_\e(x)\left(\D+\l_{\e,K}\right)
\t\psi_{\e,K}^{mid}(\xi),
\\
f^{(3)}_{\e,K}(x)&=\t\psi_{\e,K}^{mid}\D\chi(1-r)+
2\left(\nabla_x\chi(1-r),\nabla_x\t\psi_{\e,K}^{mid}(\xi)\right),
\\
f^{(4)}_{\e,K}(x)&=\sum\limits_{m=0}^{N-1}
\chi\left(\left|\vs^m\right|\eta^{1/2}\right)
\left(\D+\l_{\e,K}\right)\psi_{\e,K}^{in}\left(\vs^m\right),
\\
f^{(5)}_{\e,K}(x)&=\sum\limits_{m=0}^{N-1}\D
\chi\left(\left|\vs^m\right|\eta^{1/2}\right)\psi_{\e,K}^{mat}(x)
+2\left(\nabla_x\chi\left(\left|\vs^m\right|\eta^{1/2}\right)
,\nabla_x\psi_{\e,K}^{mat}(x)\right),
\\
\t\psi_{\e,K}^{mid}&=\psi_{\e,K}^{mid}+\e^{K+1}\a_{K+1}X,
\\
\psi_{\e,K}^{mat}&=\psi_{\e,K}^{in}-\psi_{\e,K}^{ex}-
\t\psi_{\e,K}^{mid}+\left(\e\mathsf{b}_k(\e)\xi_2+\e^{K+1}
G^{(K)}_\e+\e^{K+1}\t
v_{K+1}(0,\mu)-\e^K\frac{\a_{K+1}}{A+\mu}\right).
\end{align*}
Direct calculations yield
$\left\|f_{\e,K}^{(1)}\right\|=O\left(\e^K(A+\mu)\right)$. Due
to the equations (\ref{2.6}) the representation
\begin{equation*}
\left(\D+\l_{\e,K}\right)\t\psi_{\e,K}^{mid}(\xi)=\frac{\e^K}{r^2}
\sum\limits_{j=0}^{K-1}F^{(j)}_K(\xi,\mu),
\end{equation*}
holds, where $F^{(j)}_{K}$ are explicitly calculated functions,
and it easy to show that $F_K^{(j)}\in\mathcal{V}\cap L_2(\Pi)$,
$\left\|F^{(j)}_K\right\|=O((A+\mu))$ as $\mu\to0$, from what it
follows that
$\left\|f_{\e,K}^{(2)}\right\|=O\left(\e^{K+1/2}(A+\mu)\right)$.
By exponential decaying as $\xi_2\to+\infty$ of the functions
$v_i$ one can deduce that
$\left\|f_{\e,K}^{(3)}\right\|=O\left(\mathrm{e}^{-1/\e^{q}}
(A+\mu)\right)$, where $q$ is a some fixed number. Bearing in
mind the problems for the functions $w_{i,j}$, we see that
\begin{equation*}
\D\psi_{\e,K}^{in}=\frac{1}{r^2}\sum\limits_{i=0}^K\e^i
\left(\sum\limits_{k=0}^1\sum\limits_{j=1}^2\mathsf{a}_{kj}
\vs_2^{k+j-1}\frac{\p^j}{\p\vs_2^j}w_{k,j}+\e\eta\vs_2
\frac{\p}{\p\vs_2}\left(\vs_2\frac{\p}{\p\vs_2}\right)w_{i,1}
\right).
\end{equation*}
Using the explicit formulae for the functions $w_{i,j}$ and the
asymptotics (\ref{2.28}), we obtain the equality
$\left\|f_{\e,K}^{(4)}\right\|=O\left(\e^{1/2}\eta^{1/2}
(A+\mu)\right)$. In view of the matching carried out for
$\eta<(\xi_1-\pi m)^2+\xi_2^2<4\eta$
\begin{equation}\label{2.36}
\psi_{\e,K}^{mat}(x)=O\left(\e(A+\mu)\rho^2\ln\rho\right),
\end{equation}
from what it follows that
$\left\|f_{\e,K}^{(5)}\right\|=O\left(\eta^{1/5}\right)$.
Observe that it is impossible to get (\ref{2.36}) without
introducing the functions $w_{i,1}$, i.e., it is impossible to
attain the rapid decaying of the norm
$\left\|f_{\e,K}^{(5)}\right\|$ as $\e\to0$. This is the only
reason for that the functions $w_{i,1}$ were employed.
Collecting now the estimates for the functions $f_{\e,K}^{(i)}$,
we arrive at the desired estimate for $\left\|f_{\e,K}\right\|$.
The proof is complete.

\centsect{Formal construction of the asymptotics for the case
$n>0$}

In present section we shall formally construct the asymptotics
for the eigenvalue $\l_\e$, converging to a root $\l_0$ of the
equation (\ref{1.4}) with $n>0$, and we shall formally construct
the asymptotics for the associated eigenfunctions $\psi_\e^\pm$.

On the whole, the scheme of construction is similar to the case
$n=0$. The only (and not principal) distinction is the using of
the multiscaled method.

The asymptotics for the eigenvalue is constructed in the form of
the series (\ref{1.5}), and we construct the asymptotics of the
eigenfunctions $\psi_\e^\pm$ as the series
\begin{align}
{}&
\begin{aligned}
\psi_\e^\pm(x)=&\left(\psi^{ex,\pm}_\e(x)+\chi(1-r)
\psi^{mid,\pm}_\e(\xi,\th)\right)\chi_\e(x)+\\
{}&+\sum\limits_{m=0}^{N-1}\chi\left(\left|\vs^m\right|\eta^{1/2}
\right)\psi^{in,\pm}_\e\left(\vs^m,\mu\right),
\end{aligned}\label{3.1}
\\
{}&\psi^{ex,\pm}_\e(x)=J_n\left(\sqrt{\l_\e}r\right)\phi^\pm(n\th),
\nonumber
\\
{}&\psi^{mid,\pm}_\e(\xi,\theta)=\phi^\pm(n\theta)
\sum_{i=1}^\infty\e^i v_i(\xi,\mu)
\pm\phi^\mp(n\theta)\e\sum_{i=1}^\infty\e^i v_i^{ad}(\xi,\mu),
\label{3.3}
\\
{}&
\begin{aligned}
\psi^{in,\pm}_\e(\vs,\theta)=&\phi^\pm(n\theta)
\sum_{i=1}^\infty\e^i\left(w_{i,0}(\vs,\mu)+\e\eta
w_{i,1}(\vs,\mu)\right)\pm
\\
{}&\pm\phi^\mp(n\theta)\e \eta\sum_{i=0}^\infty \e^i
w^{ad}_{i,1}(\vs,\mu).
\end{aligned}\label{3.4}
\end{align}

By analogy with the previous section, the functions
$\psi_\e^{mid,\pm}$ are the boundary layers, we introduce them
in order to attain the Neumann boundary condition on $\G_\e$.
Employing the method of matched asymptotics expansions, we
construct the asymptotics for the eigenfunctions $\psi_\e^\pm$
in the form of the series  (\ref{3.4}) in the vicinity of the
points $x_m$ what allows us to get homogeneous Dirichlet
boundary condition on $\g_\e$. Here the distinction from the
case $n=0$ is the appearance of the additional functions
$v_i^{ad}$ and $w_{i,1}^{ad}$, and the using of multiscaled
method. To the latter it corresponds the presence of the
functions $\phi^\pm(n\th)$ in (\ref{3.3}), (\ref{3.4}), the
variable $\th$ plays the role of ''slow time''.

The objective of this section is to determine the functions
$\L_i$, $v_i$, $v_i^{ad}$, $w_{i,j}$, $w_{i,j}^{ad}$, for which
we shall obtain the explicit formulae.

We proceed to the construction. We postulate the sum of the
functions $\psi_\e^{ex,\pm}$ and $\psi_\e^{mid,\pm}$ to satisfy
homogeneous Neumann boundary condition everywhere on $\p D$
except the points  $x_k$, i.e.,
\begin{equation*}
\sqrt{\l_\e} J'_n\left(\sqrt{\l_\e}\right)\phi^\pm(n\th)-
\frac{1}{\e}\frac{\p}{\p\xi_2}\psi_\e^{mid,\pm}(\xi,\th)=0,\quad
\xi\in\G^0.
\end{equation*}
Replacing now $\l_\e$ and $\psi_\e^{mid,\pm}$ by the series
(\ref{1.5}) and (\ref{3.3}), and calculating the coefficients of
the powers of  $\e$ separately for $\phi^+(n\th)$ and
$\phi^-(n\th)$, we get the boundary conditions for the functions
$v_i^\pm$:
\begin{align}
{}&\frac{\p v_i}{\p\xi_2}=\a_i,\quad \frac{\p
v_i^{ad}}{\p\xi_2}=0, \quad \xi\in\G^0,\qquad
\a_i=\a_i(\L_0,\ldots,\L_{i-1}),\label{3.5}
\\
{}&\ \begin{aligned}
{}&\a_1=\sqrt{\L_0}J'_n\left(\sqrt{\L_0}\right),\qquad\a_2=\a_3=0,
\\
{}&\a_i=-\frac{J_n\left(\sqrt{\l_0}\right)\left(\L_0-n^2\right)}
{2\L_0}\L_{i-1}+f_i,\quad i\geq4,\qquad f_4=f_5=0,
\end{aligned}\label{3.6}
\end{align}
where $f_i=f_i(\L_0,\ldots,\L_{i-4})$ are polynomials
on$\L_1,\ldots,\L_{i-4}$ with holomorphic on $\L_0$
coefficients, $f_i(\L_0,0,\ldots,0)=0$. Similarly to the way by
which the equations (\ref{2.6}) were obtained, we substitute
(\ref{1.5}) and (\ref{3.3}) in (\ref{1.1}) and calculate the
coefficients of powers of  $\e$ separately for $\phi^+(n\th)$
and $\phi^-(n\th)$. As a result, we deduce the equations for
$v_i^\pm$:
\begin{equation}
\begin{aligned}
\D_\xi v_i=&F_i\equiv\mathcal{L}_i(v_1,\ldots,v_{i-1})-n^2
v_{i-2}+2n\frac{\p v_{i-2}^{ad}}{\p\xi_1} ,\quad\xi_2>0,
\\
\D_\xi
v_i^{ad}=&F_i^{ad}\equiv\mathcal{L}_i(v_1^{ad},\ldots,v_{i-1}^{ad})-n^2
v_{i-2}^{ad}-2n\frac{\p v_i}{\p\xi_1} ,\quad\xi_2>0,
\end{aligned}\label{3.7}
\end{equation}
where $\L_1=\L_2=0$, $v_{-1}^\pm=v_0^\pm=0$. We seek the
exponentially decaying as $\xi_2\to+\infty$ solutions of the
recurrence system of boundary value problems (\ref{3.5}),
(\ref{3.7}).

By analogy with Lemma~\ref{lm2.2} one can prove the following
statement.

\begin{lemma}\label{lm3.1}  Let the function
$F(\xi)\in\mathcal{V}^-$ has infinitely differentiable
asymptotics
\begin{equation*}
F(\xi)=\a\rho^{-1}\cos\vt+O(\ln\rho),\quad\rho\to0,
\end{equation*}
and there exists a natural number $k$, such that $\D_\xi^k
F\equiv0$ for $\xi_2>0$. Then the function $v$ defined in
accordance with (\ref{2.7}) is a solution of the boundary value
problem (\ref{2.8}), belonging to $H^1(\Pi)\cap\mathcal{V}^-$,
and having infinitely differentiable asymptotics
\begin{equation*}
v(\xi)=\frac{1}{2}\xi_1\ln\rho+\t\a\xi_1+O(\rho^2\ln\rho),
\quad\rho\to0,
\end{equation*}
where $\t\a$ is a some number.
\end{lemma}
Employing Lemmas~\ref{lm2.2} and \ref{lm3.1}, by analogy with
Lemma~\ref{lm2.3}, it is easy to prove the following lemma.

\begin{lemma}\label{lm3.2}
For each sequence $\left\{\L_i(\mu)\right\}_{i=0}^\infty$,
$\L_1(\mu)=\L_2(\mu)=0$, there exist solutions of the boundary
value problems (\ref{3.5}), (\ref{3.7}) defined by formula
(\ref{2.7}) with $F=F_i$, $k=k_i$ and $F=F^{ad}$, $k=k_i^{ad}$,
where $k_i$, $k_i^{ad}$ are some natural numbers. For the
functions $v_i$  and $v_i^{ad}$ the representations (\ref{2.17})
with $\a_i$ from (\ref{3.6}) and
\begin{equation*}
v_i^{ad}(\xi,\mu)=\sum\limits_{j=1}^{M_i^{ad}}
a_{ij}^{ad}\left(\a_1,\L_0,\ldots,\L_{i-2}\right)
v_{ij}^{ad}(\xi),
\end{equation*}
hold, where $v_{ij}\in H^1(\Pi)\cap\mathcal{V}^+$,
$v_{ij}^{ad}\in H^1(\Pi)\cap\mathcal{V}^-$, $a_{ij}$,
$a_{ij}^{ad}$ are polynomials on $\L_1,\ldots,\L_{i-2}$ with
holomorphic on $\L_0$ and $\a_1$ coefficients,
$a_{ij}\left(0,\L_0,0,\ldots,0\right)=
a_{ij}^{ad}\left(0,\L_0,0,\ldots,0\right)=0$. The equalities
$M_1$=0, $M_2=M_1^{ad}=M_2^{ad}=1$, $M_3=2$, $M_4=3$,
\begin{equation}\label{3.10}
\begin{aligned}
{}& a_{21}=a_{31}=a_{41}=a^{ad}_{11}=a^{ad}_{21}=-\a_1,\quad
a_{32}=-\a_1\left(\L_0+2n^2\right),
\\
{}& a_{42}=\frac{\a_1\left(6\L_0+1\right)}{24},\quad
a_{43}=\frac{\a_1\left(8\L_0+1\right)}{32},\quad
v_{21}=\frac{1}{2}\xi_2^2\frac{\p X}{\p\xi_2},
\\
{}& v_{11}^{ad}=-n\mathcal{A}_1\left[\frac{\p
X}{\p\xi_1}\right],\quad v_{31}=\frac{1}{8}\xi_2^4\frac{\p^4
X}{\p\xi_2^2}+\frac{1}{6}\xi_2^3\frac{\p X}{\p\xi_2}+
\frac{n^2}{2}\xi_2^2 X,
\\
{}& v_{32}=\frac{1}{2}\mathcal{A}_1[X],\quad
v_{41}=\frac{1}{48}\xi_2^6\frac{\p^3 X}{\p\xi_2}+
\frac{2}{15}\xi_2^5\frac{\p^2 X}{\p\xi_2^2}+
\frac{4n^2+1}{16}\xi_2^4\frac{\p X}{\p\xi_2},
\\
{}& v_{42}=\xi_2^3 X,\quad v_{43}=\mathcal{A}_1[\xi_2
X]+\xi_2^2\int\limits_{\xi_2}^{+\infty}X(\xi_1,t)\,dt,\quad
v_{22}^{ad}=\frac{n}{2}\xi_2^3\frac{\p X}{\p\xi_1}.
\end{aligned}
\end{equation}
take place. The asymptotics (\ref{2.19}) with $\a_i$ from
(\ref{3.6}) and
\begin{equation}\label{3.11}
v_i^{ad}(\xi,\mu)=-n\a_i\xi_1\ln\rho+\t\a_i\xi_1+
O\left(\rho^2\ln\rho\right),
\end{equation}
are correct as $\rho\to0$. Here $\a_0=0$,
$\t\a_i=\t\a_i\left(\a_1,\L_0,\ldots,\L_{i-2}\right)$ are
polynomials on $\L_1,\ldots,\L_{i-2}$ with holomorphic on $\L_0$
and $\a_1$ coefficients, moreover,
$\t\a_i\left(0,\L_0,0,\ldots,0\right)=0$.
\end{lemma}

Similarly to the previous section, for construction of the
asymptotics for the eigenfunctions $\psi_\e^\pm$ in a
neighbourhood of the points $x_m$ we apply the method of matched
asymptotics expansions. The asymptotics of the functions
$\psi_\e^\pm$ in a neighbourhood of the points $x_m$ are
constructed in the form of the series (\ref{3.4}), doing this,
we match the functions $w_{i,j}$ with $v_i$, whereas the
functions $w_{i,1}^{ad}$ are matched with $v_i^{ad}$.

We substitute the series (\ref{1.5}) and (\ref{3.4}) in the
problem (\ref{1.1}), (\ref{1.2}), pass to the variables  $\vs$
and collect coefficients of powers of $\e$ separately for
$\phi^+(n\th)$ and $\phi^-(n\th)$. As a result, we get the
boundary value problems (\ref{2.20}) and (\ref{2.21}) for the
functions $w_{i,j}$, and the following ones for the functions
$w_{i,1}^{ad}$:
\begin{equation}\label{3.12}
\begin{aligned}
\D_\vs w_{i,1}^{ad}&=2n\frac{\p}{\p\vs_1}w_{i,0},\quad\vs_2>0,
\\
w_{i,1}^{ad}&=0,\quad\vs\in\g^1,\qquad
\frac{\p}{\p\vs_2}w_{i,1}^{ad}=0,\quad \vs\in\G^1,
\end{aligned}
\end{equation}
where $w_{0,0}=0$. Let us deduce the asymptotics for $w_{i,j}$
and $w_{i,1}^{ad}$ as $|\vs|\to\infty$. We denote by $\l_{\e,K}$
partial sum of (\ref{1.5}),
\begin{align*}
{}&\Psi_{\e,K}^\pm(x)=\psi_{\e,K}^{ex,\pm}(x)+
\chi(1-r)\psi_{\e,K}^{mid,\pm}(\xi),\quad\psi_{\e,K}^{ex,\pm}(x)=
J_n\left(\sqrt{\l_{\e,K}}r\right)\phi^\pm(n\th),
\\
{}&
\psi_{\e,K}^{mid,\pm}(\xi)=\phi^\pm(n\th)\sum\limits_{i=1}^{K+1}\e^i
v_i(\xi,\mu)\pm\phi^\mp(n\th)\e\sum\limits_{i=1}^{K}\e^i
v_i^{ad}(\xi,\mu).
\end{align*}
It is easily seen that
\begin{align}
{}&\begin{aligned}
J_n\left(\sqrt{\l_{\e,K}}\right)=&\sum\limits_{i=0}^{K}\e^i
G_i(\L_0,\ldots,\L_i)+\e^{K+1}G^{(K)}_\e\left(\L_0,\ldots,
\L_K\right)-\\&-\e\xi_2\sqrt{\l_{\e,K}}J'_n\left(\sqrt{\l_{\e,K}}\right)
+O\left(\e^2\xi^2_2\right),
\end{aligned}\nonumber
\\
{}&\qquad
\begin{aligned}
{}&G_0=J_n\left(\sqrt{\L_0}\right),\qquad G_1=G_2=0,
\\
{}&G_i=\frac{ J'_n\left(\sqrt{\L_0}\right)}{2\sqrt{\L_0}}\L_i+
g_i,\quad i\ge3,\qquad g_3=g_4=0,
\end{aligned}\label{3.13}
\end{align}
where the functions $g_i=g_i\left(\L_0,\ldots,\L_{i-3}\right)$
are polynomials on $\L_1,\ldots,\L_{i-3}$ with holomorphic on
$\L_0$ coefficients, $g_i\left(\L_0,0,\ldots,0\right)=0$,
$G^{(K)}_\e$ is a bounded holomorphic on  $\L_1,\ldots,\L_K$
function, $G^{(K)}_\e\left(\L_0,0,\ldots,0\right)=0$. From the
relations obtained, the asymptotics (\ref{2.19}) and
(\ref{3.11}), and the equality $\ln\eta=-\frac{1}{\e(A+\mu)}$ it
follows that
\begin{align*}
\Psi_{\e,M}^\pm(x)&=\phi^\pm(n\th)\left(\sum\limits_{i=0}^{K}\e^i
W_{i,0}(\vs,\mu)+\e \eta\sum\limits_{i=1}^K\e^i
W_{i,1}(\vs,\mu)+W^{(K)}_\e(x)\right)\pm
\\
{}&\pm\phi^\mp(n\th)\e\eta\left(\sum\limits_{i=0}^K\e^i
W_{i,1}^{ad}(\vs,\mu)+\e^{K+1}
W_K^{ad}(\vs,\mu)\right)+O\left(\e\eta^2|\vs|^2\ln|\vs|\right),
\end{align*}
as $\eta^{1/2}<\rho<2\eta^{1/2}$, where $W_{i,j}$,
$W^{(K)}_\e(x)$ from (\ref{2.26}), (\ref{2.27}) with $\a_i$ from
(\ref{3.6}),
\begin{align}
W_{i,1}^{ad}(\vs,\mu)&=-n\a_i\vs_1\ln|\vs|+\left(\t\a_i+
\frac{n\a_{i+1}}{A+\mu}\right)\vs_1,\label{3.14}
\\
W_K^{ad}(\vs,\mu)&=-\frac{n\a_{K+1}}{A+\mu}\vs_1,\nonumber
\\
W^{(K)}_\e(x)&=\e^{K+1}\left(-\a_{K+1}\left(\ln|\vs|+\ln
2-\xi_2\right)+G^{(K)}_\e+\t v_{K+1}(0,\mu)\right)+
\e\mathsf{b}_K(\e)\xi_2,\nonumber
\\
\mathsf{b}_K(\e)&=\sum\limits_{i=1}^K\e^{i-1}\a_i-
\sqrt{\l_{\e,K}}J'_n\left(\sqrt{\l_{\e,K}}\right),\nonumber
\end{align}
$\t\a_0=0$. Following the method of matched asymptotics
expansions, we must construct the solutions of the problems
(\ref{2.20}), (\ref{2.21}) and (\ref{3.12}) with asymptotics
(\ref{2.28}) and
\begin{equation}\label{3.16}
w^{ad}_{i,1}(\vs,\mu)=W_{i,1}^{ad}(\vs,\mu)+o(|\vs|),\quad
|\vs|\to\infty.
\end{equation}
We define the functions $w_{i,j}$ in accordance with
(\ref{2.30}) and (\ref{2.33}), where $\a_i$ from (\ref{3.6}).
Then in view of the definition of the function $w_{i,0}$ and the
asymptotics (\ref{2.26}), (\ref{2.28}) we deduce the equality
(\ref{2.31}), where $\a_i$ from (\ref{3.6}), $\t v_i(0,\mu)$
from Lemma~\ref{lm3.2}, $G_i$ from (\ref{3.13}). For $i=0$, this
equality becomes the equation (\ref{1.6}) for $\L_0$. Since for
$n>0$ the eigenvalue $\l_0$ is nonzero, the holomorphy of $\L_0$
easily follows from the implicit function theorem. The
equalities (\ref{2.31}) hold for $i=1,2$, since by (\ref{3.6}),
(\ref{3.10}), and (\ref{3.13}) we have $\a_2=\a_3=G_1=G_2=0$,
$\t v_1(0,\mu)=\t v_2(0,\mu)=0$. For $i\ge3$, by analogy with
the way by which (\ref{2.32}) was obtained from (\ref{1.6}),
(\ref{2.31}), (\ref{3.6}) and (\ref{3.13}), one can get the
formulae for $\L_i$:
\begin{align*}
{}&\L_i(\mu)=\frac{2\L_0(\mu)\left(\t f_{i+1}(\mu)+(A+\mu)
\left(\t g_i(\mu)+\t
v_i(0,\mu)\right)\right)}{J_n(\sqrt{\L_0(\mu)})
\left(\L_0(\mu)-n^2+(A+\mu)^2\right)},
\\
{}& \t f_{i+1}(\mu)=
f_{i+1}\left(\L_0(\mu),\ldots,\L_{i-3}(\mu)\right), \quad \t
g_i(\mu)=g_i\left(\L_0(\mu),\ldots,\L_{i-3}(\mu)\right).
\end{align*}
Making $i=3,4$ in the formulae obtained and using the equalities
(\ref{2.17}), (\ref{2.37}) and (\ref{3.10}), we have (\ref{1.8})
also for $n>0$. Reproducing the arguments of the previous
section, one can prove that $\L_i(\mu)$ are holomorphic on
$\mu\ge0$ functions satisfying the representations
$\L_i(\mu)=\mu^2\t\L_i(\mu)$ for $A=0$, where $\t\L_i(\mu)$ are
holomorphic functions. Let us construct the functions
$w_{i,1}^{ad}$. It is easy to see, that the functions
\begin{equation*}
Y_1(\vs)=\mathrm{Re}\,\sqrt{y^2-1},\qquad
Y_2(\vs)=\frac{1}{2}\left(\vs_1 Y(\vs)-\ln 2 \,Y_1(\vs)\right)
\end{equation*}
are solutions of the boundary value problems
\begin{align*}
{}&\D_\vs Y_1=0,\quad \D_\vs Y_2=\frac{\p
Y}{\p\vs_1},\quad\vs_2>0,
\\
{}&Y_j=0,\quad \vs\in\g^1,\qquad\frac{\p
Y_j}{\p\vs_2}=0,\quad\vs\in\G^1,\quad j=1,2,
\end{align*}
and have asymptotics
\begin{equation*}
Y_1(\vs)=\vs_1+O(|\vs|^{-1}),\qquad
Y_2(\vs)=\frac{1}{2}\vs_1\ln|\vs|+
O(|\vs|^{-1}),\quad|\vs|\to\infty.
\end{equation*}
By the properties of the functions $Y_j$, the definition of the
function $w_{i,0}$, the problem (\ref{3.12}) and the asymptotics
(\ref{3.14}), (\ref{3.16}) we obtain that
\begin{equation*}
w_{i,1}^{ad}=-2n\a_i
Y_2+\left(\t\a_i+\frac{\a_{i+1}}{A+\mu}\right)Y_1.
\end{equation*}
We set
\begin{align*}
\psi_{\e,K}^\pm(x)=&\Psi_{\e,K}^\pm(x)\chi_\e(x)+
\sum\limits_{m=0}^{N-1}\chi\left(\left|\vs^m\right|\eta^{1/2}
\right)\psi^{in,\pm}_{\e,K}\left(\vs^m\right),
\\
\psi^{in,\pm}_{\e,K}\left(\vs\right)=&\phi^\pm(n\th)\sum_{i=1}^K\e^i\left(
w_{i,0}(\vs,\mu) +\e\eta w_{i,1}(\vs,\mu)\right)\pm
\\
\pm&\phi^\mp(n\th)\e\eta\sum\limits_{i=0}^K\e^i
w_{i,1}^{ad}(\vs,\mu),
\\
\widetilde{\psi}_{\e,K}^\pm(x)=&\psi_{\e,K}^\pm(x)-
R^\pm_{\e,K}(x),
\\
R^\pm_{\e,K}(x)=&\chi_\e(x)\phi^\pm(n\th)\t R_{\e,K}(x)\pm
\chi_\e(x)\phi^\mp(n\th)\frac{n\a_{K+1}}{A+\mu}\sum\limits_{m=0}^{N-1}
\chi(|\vs^m|\eta^{1/2})Y_1(\vs^m),
\\
\t R_{\e,K}(x)=&\chi(1-r)
\left(\mathsf{b}_k(\e)(1-r)-\e^{K+1}\a_{K+1}X(\xi)\right)+
\\
{}&+\e^{K+1}G^{(K)}_\e+\e^{K+1}\t
v_{K+1}(0,\mu)-\e^K\frac{\a_{K+1}}{A+\mu}.
\end{align*}
By analogy with Theorem~\ref{th2.1}, one can prove the following
statement.
\begin{theorem}\label{th3.1}
The functions $\psi_{\e,K}^\pm, \t\psi_{\e,K}^\pm\in H^1(D)\cap
C^\infty(D)$ converge to $\psi_0^\pm$ in $L_2(D)$ as $\e\to0$,
$\l_{\e,K}$ converges to $\l_0$,
$\left\|R_{\e,K}^\pm\right\|=O(\e^{K}(A+\mu))$.The functions
$\t\psi_{\e,K}^\pm$ and $\l_{\e,k}$ are the solutions of the
problem (\ref{2.35}) with $u_\e=\t\psi_{\e,K}^\pm$,
$\l=\l_{\e,K}$, $f=f_{\e,K}^\pm$, where
$\left\|f_{\e,K}^\pm\right\|=O\left(\e^K(A+\mu)\right)$.
\end{theorem}

\centsect{Justification of the asymptotics}

In this section we shall prove that asymptotic expansions
formally constructed in two previous sections are really provide
asymptotics for the eigenelements of the problem (\ref{1.1}),
(\ref{1.2}). In order to do it we shall employ the following
statements.
\begin{lemma}\label{lm4.1}
Let $Q$ be any compact set in complex plane containing no
eigenvalues of the limiting problem. Then for all $\l\in Q$,
$f\in L_2(D)$ and sufficiently small  $\e$ the problem
(\ref{2.35}) is uniquely solvable and for its solution the
uniform on  $\e$, $\mu$, $\l$ and $f$ estimate
\begin{equation}
\|u_\e\|_1\le C\|f\|,\label{4.1}
\end{equation}
holds, where $\|\bullet\|_1$ is the $H^1(D)$-norm. The function
$u_\e$ converges to the solution of the problem
\begin{equation}\label{4.6}
-\D u_0=\l u_0+f,\quad x\in D,\qquad \left(\frac{\p}{\p r
}+A\right)u_0=0,\quad x\in\p D.
\end{equation}
uniformly on $\l$.
\end{lemma}

\noindent\textbf{Proof.} The solvability of the problem
(\ref{2.35}) is obvious. Clear, in order to prove the uniqueness
of its solution it is sufficient to prove the estimate
(\ref{4.1}). We prove the latter by arguing by contradiction.
Suppose that there exist sequences
$\e_k\underset{k\to\infty}{\longrightarrow}0$, $f_k$ and $\l_k$
such that for $\e=\e_k$, $f=f_k$, $\l=\l_k\in Q$ the inequality
\begin{equation}\label{4.2}
\left\|u_{\e_k}\right\|_1\ge k\|f_k\|.
\end{equation}
takes place. There is no loss of generality in assuming that
$\|u_\e\|=1$. We multiply both sides of the equation in
(\ref{2.35}) by $u_\e$ and integrate by part. Then we have a
priori uniform estimate
\begin{equation*}
\|u_\e\|_1\le C\left(\|f\|+\|u_\e\|\right).
\end{equation*}
By this estimate, the equality $\|u_\e\|=1$, and (\ref{4.2}) we
deduce
\begin{equation}\label{4.4}
\left\|u_{\e_k}\right\|_1\le C,\qquad \left\|f_k\right\|
\underset{k\to\infty}{\longrightarrow}0.
\end{equation}
From the assertions obtained and the theorem about the compact
embedding of $H^1(D)$ in $L_2(D)$ it follows that there exists a
subsequence of indexes $k$ (we indicate it by $k'$), such that
$\l_{k'}\to\l_*\in Q$ and
\begin{equation*}
u_{\e_{k'}}\to u_*\not=0 \qquad\text{weakly in $H^1(D)$ and
strongly in $L_2(D)$}.
\end{equation*}
In \ref{Ch} it was shown that for each function $V\in
C^\infty(\overline{D})$ there exists a sequence of functions
$V_\e\in H^1(D)$, vanishing on $\g_\e$, such that
\begin{equation}\label{4.5}
\begin{aligned}
{}&V_\e\to V\qquad\text{weakly in $H^1(D)$ and strongly in
$L_2(D)$},
\\
{}&\int\limits_D\left(\nabla V_\e,\nabla v_\e\right)\,dx\to
\int\limits_D\left(\nabla V,\nabla v\right)\,dx+A
\int\limits_{\p D} V v\,d\th,
\end{aligned}
\end{equation}
where $v_\e$ is an arbitrary sequence of functions from
$H^1(D)$, $v_\e=0$ on $\g_\e$, $v_\e$ converges to $v\in H^1(D)$
strongly in $L_2(D)$ and weakly in $H^1(D)$. In view of
(\ref{2.35}) we have the equality
\begin{equation*}
\int\limits_D\left(\nabla V_{\e_{k'}},\nabla
u_{\e_{k'}}\right)\,dx=\l_{k'}\int\limits_D V_{\e_{k'}}
u_{\e_{k'}}\,dx+\int\limits_D f_{k'} u_{\e_{k'}}\,dx,
\end{equation*}
passing in which to limit as $k'\to\infty$ and bearing in mind
(\ref{4.4}), (\ref{4.5}), we conclude that $u_*$ is a solution
of the problem
\begin{equation*}
-\D u_*=\l u_*,\quad x\in D,\qquad \left(\frac{\p}{\p r
}+A\right)u_*=0,\quad x\in\p D,
\end{equation*}
i.e., $\l_*\in Q$ is an eigenvalue of the limiting problem,
whereas by assumption the set $Q$ does not contain the
eigenvalues of the limiting problem, a contradiction. The
estimate (\ref{4.1}) is proved.

By similar arguments, employing (\ref{4.1}) instead of
(\ref{4.4}), it easy to prove the convergence of the solution of
the problem (\ref{2.35}) with
$\l=\l(k)\underset{k\to\infty}{\longrightarrow}\l_*$ to the
solution of the problem (\ref{4.6}) with $\l=\l_*$. From this
fact and continuity on $\l$ of  $u_0$ it follows the uniform on
$\l$ convergence of $u_\e$ to $u_0$. The proof is complete.

\begin{lemma}\label{lm4.2}
Let $\l_0$ be a $p$-multiply eigenvalue of the limiting problem,
$\l_\e^{(j)}$, $j=1,\ldots,p $ be the eigenvalues of the
perturbed problem, converging to $\l_0$, with multiplicity taken
into account, $\psi_\e^{(j)}$ be the associated eigenfunctions
orthonormalized in $L_2(D)$. Then for $\l$ close to $\l_0$ for
solution of the problem  (\ref{2.35}) the representation
\begin{equation}\label{4.7}
u_\e=\sum\limits_{j=1}^p\frac{\psi_\e^{(j)}}{\l_\e^{(j)}-\l}
\int\limits_D\psi_\e^{(j)}f\,dx+\t u_\e,
\end{equation}
holds, where $\t u_\e$ is a holomorphic (in $L_2(D)$-norm) on
$\l$ function, orthogonal in $L_2(D)$ to all $\psi_\e^{(j)}$.
For $u_\e$ uniform on $\e$, $\mu$, $\l$ and $f$ estimate
\begin{equation}\label{4.8}
\left\|\t u_\e\right\|_1\le C\|f\|.
\end{equation}
takes place.
\end{lemma}

\noindent\textbf{Proof.} It is known that the solution $u_\e$ of
the problem (\ref{2.35}) is a meromorphic on $\l$ function
having only simple poles coinciding with the eigenvalues of the
perturbed problem. Residua at these poles (eigenvalues) are the
associated eigenfunctions of the perturbed problem. Since
$\l_\e^{(j)}$ converge to $\l_0$, then $\l$, close to $\l_0$,
are close to $\l_\e^{(j)}$. For this reason, for the function
$u_\e$ the representation
\begin{equation}\label{4.9}
u_\e=\sum\limits_{j=1}^p
\frac{b^{(j)}_\e}{\l_\e^{(j)}-\l}\psi_\e^{(j)}+\t u_\e,
\end{equation}
is valid, where $b^{(j)}_\e$ are some scalar coefficients, $\t
u_\e$ is a holomorphic on $\l$ function. From the equation for
$u_\e$ it follows that
\begin{equation*}
\left(\l_\e^{(j)}-\l\right)\int\limits_D\psi_\e^{(j)}u_\e\,dx=
\int\limits_D\psi_\e^{(j)}f\,dx.
\end{equation*}
Substituting the formula (\ref{4.9}) into this equality we
obtain that
\begin{equation*}
b^{(j)}_\e+\left(\l_\e^{(j)}-\l\right)
\int\limits_D\psi_\e^{(j)}\t u_\e\,dx=
\int\limits_D\psi_\e^{(j)}f\,dx,
\end{equation*}
from what by holomorphy of $\t u_\e$ we deduce:
\begin{equation*}
b^{(j)}_\e= \int\limits_D\psi_\e^{(j)}f\,dx,\quad
\int\limits_D\psi_\e^{(j)}\t u_\e\,dx=0,\quad j=1,\ldots,p .
\end{equation*}
These relations and  (\ref{4.9}) imply (\ref{4.7}).

Let us show the estimate (\ref{4.8}). We indicate by $S(z,a)$ an
open circle  of radius $a$ in complex plane with center at the
point  $z$. We choose the number $\d$ by the condition that the
circle $S(\l_0,\d)$ contains no eigenvalues of the limiting
problem except $\l_0$. Then for all sufficiently small $\e$ each
$\l_\e^{(j)}$ lies in the circle $S(\l_0,\d/2)$. Therefore, by
the representation (\ref{4.9}) and Lemma~\ref{lm4.1} for
$\l\in\p S(\l_0,\d)$ the uniform estimate
\begin{equation*}
\left\|\t u_\e\right\|_1=\left\|u_\e-
\sum\limits_{j=1}^p\frac{\psi_\e^{(j)}}{\l_\e^{(j)}-\l}
\int\limits_D\psi_\e^{(j)}f\,dx\right\|_1\le
C\|f\|+\frac{2}{\d}\sum\limits_{j=1}^p\left\|\psi_\e^{(j)}
\right\|_1\|f\|\le C\|f\|.
\end{equation*}
is true.  Since $\t u_\e$ is holomorphic on $\l$, then due to
module maximum principle the last inequality holds also for
$\l\in S(\l_0,\d)$. The proof is complete.

\begin{lemma}\label{lm4.3}
The eigenvalues of the perturbed problem have the asymptotics
(\ref{1.5})--(\ref{1.8}).
\end{lemma}

\noindent\textbf{Proof.} Let $\l_0$ be an eigenvalue of the
problem (\ref{1.3}) and be a root of the equation (\ref{1.4})
for $n=n_i$, $i=1,\ldots,m $, where $n_i$ are different. We
suppose that $n_1=0$, $n_i>0$, $i=2,\ldots,m $. The cases
$n_i>0$, $i=1,\ldots,m $, and $m=1$, $n_1=0$, are proved in the
similar way. The eigenfunctions associated with $\l_0$ have the
form
\begin{align*}
\psi_0^{(1)}(x)&=J_0\left(\sqrt{\l_0}r\right),
\\
\psi_0^{(2i-2)}(x)&=J_{n_i}\left(\sqrt{\l_0}r\right)\phi^+\left(n_i\th\right),
\quad i=2,\ldots,m ,
\\
\psi_0^{(2i-1)}(x)&=J_{n_i}\left(\sqrt{\l_0}r\right)\phi^-\left(n_i\th\right),
\quad i=2,\ldots,m .
\end{align*}
Similarly, we denote by $\psi_{\e,K}^{(j)}$,
$\t\psi_{\e,K}^{(j)}$, $f_{\e,K}^{(j)}$ the functions
$\psi_{\e,K}$, $\psi_{\e,K}^\pm$, $\t\psi_{\e,K}$,
$\t\psi_{\e,K}^\pm$, $f_{\e,K}$, $f_{\e,K}^\pm$, constructed in
second and third sections and associated with the indexes $n_i$.
Let $\l_{\e,K}^{(1)}=\l_{\e,K}$,  where $\l_{\e,K}$ was defined
in the second section,
$\l_{\e,K}^{(2i-2)}=\l_{\e,K}^{(2i-1)}=\l_{\e,K}$, where
$\l_{\e,K}$ was defined in the third section and associated with
the index $n_i$, $i=2,\ldots,m $. Clear, the multiplicity of
$\l_0$ equals $(2m-1)$. Due to Theorems~\ref{th2.1} and
\ref{th3.1} and Lemma~\ref{lm4.2} for the functions
$\t\psi_{\e,K}^{(j)}$ the representations
\begin{align}
{}&\t\psi_{\e,K}^{(j)}=\sum\limits_{k=1}^{2m-1}\mathsf{b}_\e^{jk}
\psi_\e^{(k)}+\t u_\e^{(j)},\label{4.10}
\\
{}&\mathsf{b}_\e^{jk}=\int\limits_D\psi_\e^{(k)}\t\psi_{\e,K}^{(j)}
\,dx=\frac{1}{\l_\e^{(k)}-\l_{\e,K}^{(j)}}
\int\limits_D\psi_\e^{(k)}f_{\e,K}^{(j)}\,dx,\nonumber
\\
{}&\left\|\t u_\e^{(j)}\right\|_1\le
C\left\|f_{\e,K}^{(j)}\right\|
=O\left(\e^K(A+\mu)\right)\nonumber
\end{align}
hold. Suppose that some of the eigenvalues $\l_\e^{(j)}$ do not
satisfy the asymptotics (\ref{1.5})--(\ref{1.8}), namely,
uniform on $\e$ and $\mu$ estimate
\begin{equation}\label{4.11}
\left|\l_\e^{(k)}-\l_{\e,K}^{(j)}\right|\ge C\e^p(A+\mu),\qquad
j=1,\ldots,2m-1,\quad k\in I,
\end{equation}
hold, where $p$ \   is a some number independent on $K$, $I$ is
a subset of the indexes, $I\subseteq\{1,2,\ldots,2m-1\}$. By
(\ref{4.11}) and the statement of
Theorems~\ref{th2.1},~\ref{th3.1} for the functions
$f^{(j)}_{\e,K}$, we deduce that for $K\ge p+1$ the convergences
$\mathsf{b}_\e^{jk}\to0$, $k\in I$, $j=1,\ldots,2m-1$ hold. From
the definition of the functions $\mathsf{b}_\e^{jk}$, the
orthogonality of $\psi_\e^{(j)}$ and the convergence
$\t\psi_{\e,K}^{(j)}\to\psi_0^{(j)}$ it follows that
$\mathsf{b}_\e^{jk}$ are bounded, so, there exists a subsequence
$\e'\to0$, for that $\mathsf{b}_{\e'}^{jk}\to\mathsf{b}^{jk}_0$,
moreover, $\mathsf{b}^{jk}_0=0$, if $k\in I$, $j=1,\ldots,2m-1$.
In view of (\ref{4.10}) and Lemma~\ref{lm4.2} we have the
equalities
\begin{equation*}
\int\limits_D\t\psi_{\e',K}^{(j)}\t\psi_{\e',K}^{(l)}\,dx=
\sum\limits_{k=1}^{2m-1}\mathsf{b}_{\e'}^{jk}
\mathsf{b}_{\e'}^{lk}+ \int\limits_D\t u_{\e',K}^{(j)}
u_{\e',K}^{(l)}\,dx,
\end{equation*}
passing to limit as $\e'\to0$ in which and bearing in mind the
convergences $\t\psi_{\e',K}^{(i)}\to\psi_0^{(i)}$, and the
estimate for the functions $\t u_{\e',K}^{(i)}$, we get
\begin{equation}\label{4.12}
\mathsf{c}_{jl}\d_{jl}=\sum\limits_{k=1}^{2m-1}
\mathsf{b}_0^{jk} \mathsf{b}_0^{lk},\quad \mathsf{c}_{jj}\not=0,
\end{equation}
where $\d_{jl}$ is the Kronecker delta. Let $\mathsf{b}_0^{(j)}$
be a vector with components $\mathsf{b}_0^{jk}$,
$k=1,\ldots,2m-1$, $j\not\in I$, $j=1,\ldots,2m-1$. In view of
(\ref{4.12}) we have $(2m-1)$ nonzero orthogonal $q$-dimensional
vectors $\mathsf{b}_0^{(i)}$, where $q<2m-1$. The contradiction
obtained proves the lemma.

\begin{lemma}\label{lm4.4}
Let $\l_\e^{(1)}$ and $\l_\e^{(2)}$ be eigenvalues of the
problem (\ref{1.1}), (\ref{1.2}), having asymptotics
(\ref{1.5})--(\ref{1.8}), associated with indexes $n$ and $m$,
$n\not=m$. Then uniform on $\e$ and $\mu$ estimate
\begin{equation}\label{4.13}
\left|\l_\e^{(1)}-\l_\e^{(2)}\right|\ge C\e^4(A+\mu).
\end{equation}
holds.
\end{lemma}

\noindent\textbf{Proof.} If $\l_\e^{(i)}$, $i=1,2$, converge to
different limiting eigenvalues, then the estimate (\ref{4.13})
is obvious. So, we assume that $\l_\e^{(i)}$ converge to a same
eigenvalue $\l_0$.  First we consider the case $A=0$. Then
\begin{align*}
{}&\l_\e^{(1)}=\l_0+\mu\frac{2\l_0}{\l_0-n^2}+O(\mu(\mu+\e^3)),
\quad
\l_\e^{(2)}=\l_0+\mu\frac{2\l_0}{\l_0-m^2}+O(\mu(\mu+\e^3)),
\\
{}&\l_\e^{(1)}-\l_\e^{(2)}=\mu\frac{2\l_0(n^2-m^2)}
{(\l_0-n^2)(\l_0-m^2)}+O(\mu(\mu+\e^3)),
\end{align*}
from what it follows (\ref{4.13}) for $A=0$. We proceed to the
case $A>0$. If $\e=o(\mu^{1/3})$, then by, (\ref{1.6}), we
deduce
\begin{equation*}
\l_\e^{(1)}-\l_\e^{(2)}=\mu\frac{2\l_0(n^2-m^2)}
{(\l_0-n^2+A^2)(\l_0-m^2+A^2)}+O(\e^3),
\end{equation*}
from what it follows (\ref{4.13}) for $A>0$, $\e=o(\mu^{1/3})$.
If $\mu=o(\e^3)$, then
\begin{align*}
\l_\e^{(1)}-\l_\e^{(2)}=&-\e^3\frac{A^2\l_0\z(3)}{4}\left(
\frac{\l_0+2n^2}{\l_0-n^2+A^2}-\frac{\l_0+2m^2}{\l_0-m^2+A^2}
\right)+O(\e^4+\mu)=
\\
=&-\e^3\frac{A^2\l_0\z(3)(2A^2+3\l_0)(n^2-m^2)}
{(\l_0-n^2+A^2)(\l_0-m^2+A^2)}+O(\e^4+\mu),
\end{align*}
i.e., the estimate (\ref{4.13}) is true in this case, too. If
$\mu=O(\e^3)$, then it easy to see that
\begin{align*}
\l_\e^{(1)}-\l_\e^{(2)}=&\frac{\l_0(n^2-m^2)\left(8\mu-\e^3
A^2\z(3)(2A^2+3\l_0)\right)}{4(\l_0-n^2+A^2)(\l_0-m^2+A^2)}+
\\
{}&+\e^4\frac{\pi^4\l_0(8\l_0+1)(n^2-m^2)}
{5760(\l_0-n^2+A^2)(\l_0+2m^2-A^2)}+O(\e^5),
\end{align*}
The first term in the formula obtained being nonzero, the
inequality (\ref{4.13}) takes place. The first term being zero,
the second term does not vanish and we arrive at (\ref{4.13})
again. The proof is complete.

\noindent\textbf{Proof of Theorem~\ref{th1.1}.} Hereafter we
employ the notations introduced in the proof of
Lemma~\ref{lm4.3} and we only deal with the case considered
there (proof of other cases is similar). Let us prove that an
eigenvalue $\l_\e^{(j)}$ is simple if associated number $n_i$
equals zero and it is double if this number is positive. We
consider the eigenvalues $\l_\e^{(2p-2)}$ and $\l_\e^{(2p-1)}$
associated with the same number $n_p>0$. Due to
Lemma~\ref{lm4.3} these eigenvalues have the same asymptotic
expansions. Let us show that they are equal, too. Suppose that
they are different. In view of Lemma~\ref{lm4.4} other
eigenvalues of the perturbed problem converging to $\l_0$ have
asymptotics distinct from the asymptotics for $\l_\e^{(2p-2)}$
and $\l_\e^{(2p-1)}$. For this reason, the assumption that
$\l_\e^{(2p-2)}\not=\l_\e^{(2p-1)}$ means that they are simple.
To prove that they are coincide is to prove that the eigenvalue
$\l_\e^{(2p-1)}$ is double.  We write the representations
(\ref{4.10}) for the functions $\t\psi_{\e,K}^{(2p-2)}$ and
$\t\psi_{\e,K}^{(2p-1)}$:
\begin{equation}\label{4.14}
\begin{aligned}
\t\psi_{\e,K}^{(i)}=&\sum\limits_{j=2p-2}^{2p-1}
\mathsf{b}_\e^{ij}\psi_\e^{(j)}+\widehat{u}_\e^{(i)},\quad
i=2p-2,2p-1,
\\
\widehat{u}_\e^{(i)}=&\t u_\e^{(i)}+\sum\limits^{2m-1}_{
\genfrac{}{}{0pt}{2}{k=1}{\genfrac{}{}{0pt}{2}{k\not=2p-2}{k\not=2p-1}
} }\mathsf{b}_\e^{ik}\psi_\e^{(k)}.
\end{aligned}
\end{equation}
By Lemma~\ref{lm4.4} and the definition of the quantities
$\mathsf{b}_\e^{ik}$ we get that
$\mathsf{b}_\e^{ik}=O(\e^{K-4})$, $i=2p-2,2p-1$,
$k=1,\ldots,2m-1$, $k\not=2p-2,2p-1$. Since $\|\t u_\e^{(i)}
\|=O(\e^K(A+\mu))$, then $\|\widehat{u}_\e^{(i)}\|\to0$ as
$\e\to0$ if $K\ge5$. In view of Theorem~\ref{th3.1} we have the
convergences $\t\psi_{\e,K}^{(2p-2)}\to
J_{n_p}\left(\sqrt{\l_0}r\right)\phi^+(n_p\th)$,
$\t\psi_{\e,K}^{(2p-1)}\to
J_{n_p}\left(\sqrt{\l_0}r\right)\phi^-(n_p\th)$ as $\e\to0$. So,
there are two linear combinations of the eigenfunctions
$\psi_\e^{(2p-2)}$ and $\psi_\e^{(2p-1)}$ converging to
$J_{n_p}\left(\sqrt{\l_0}r\right)\phi^\pm(n_p\th)$:
\begin{equation}\label{4.15}
\begin{aligned}
{}&\mathsf{c}^{(1)}_\e\psi_\e^{(2p-2)}+\mathsf{c}^{(2)}_\e\psi_\e^{(2p-2)}
\to J_{n_p}\left(\sqrt{\l_0}r\right)\phi^+(n_p\th),
\\
{}&\mathsf{c}^{(3)}_\e\psi_\e^{(2p-2)}+\mathsf{c}^{(4)}_\e\psi_\e^{(2p-2)}
\to J_{n_p}\left(\sqrt{\l_0}r\right)\phi^-(n_p\th),
\end{aligned}
\end{equation}
$\mathsf{c}^{(1)}_\e=\mathsf{b}^{2p-2,2p-2}_\e$,
$\mathsf{c}^{(2)}_\e=\mathsf{b}^{2p-2, 2p-1}_\e$,
$\mathsf{c}^{(3)}_\e=\mathsf{b}^{2p-1,2p-2}_\e$,
$\mathsf{c}^{(4)}_\e=\mathsf{b}^{2p-1,2p-1}_\e$.  We introduce
the functions
$\widehat{\psi}_\e^{(i)}(r,\th)=\psi_\e^{(i)}(r,\th)\left(r,\th+
\left[\frac{N}{4n_p}\right]\e\pi\right)$, $i=2p-2,2p-1$,
$\left[\bullet\right]$ is the integral part of a number. One can
see that $\widehat{\psi}_\e^{(i)}$ are eigenfunctions of the
perturbed problem associated with $\l_\e^{(i)}$. By assumption,
the eigenvalues $\l_\e^{(i)}$, $i=2p-2,2p-1$ are simple and the
associated eigenfunctions are orthonormalized in $L_2(D)$. Thus,
$\widehat{\psi}_\e^{(2p-2)}=\mathsf{c}^{(5)}_\e\psi_\e^{(2p-2)}$,
$\widehat{\psi}_\e^{(2p-1)}=\mathsf{c}^{(6)}_\e\psi_\e^{(2p-1)}$,
$|\mathsf{c}^{(5)}_\e|=|\mathsf{c}^{(6)}_\e|=1$. From these
equalities and (\ref{4.15}) we obtain that
\begin{equation}\label{4.16}
\begin{aligned}
{}&\mathsf{c}^{(5)}_\e\mathsf{c}^{(1)}_\e\psi_\e^{(2p-2)}+\mathsf{c}^{(6)}_\e\mathsf{c}^{(2)}_\e\psi_\e^{(2p-2)}
\to -J_{n_p}\left(\sqrt{\l_0}r\right)\phi^-(n_p\th),
\\
{}&\mathsf{c}^{(5)}_\e\mathsf{c}^{(3)}_\e\psi_\e^{(2p-2)}+\mathsf{c}^{(6)}_\e\mathsf{c}^{(4)}_\e\psi_\e^{(2p-2)}
\to J_{n_p}\left(\sqrt{\l_0}r\right)\phi^+(n_p\th).
\end{aligned}
\end{equation}
Calculating scalar product (in $L_2(D)$) for the first relation
in (\ref{4.15}) and the second one in (\ref{4.16}) and for the
second relation in (\ref{4.15}) and the first one in
(\ref{4.16}), we arrive at the convergences
\begin{equation*}
\mathsf{c}^{(1)}_\e\mathsf{c}^{(3)}_\e\mathsf{c}^{(5)}_\e+
\mathsf{c}^{(2)}_\e\mathsf{c}^{(4)}_\e\mathsf{c}^{(6)}_\e\to-\mathsf{c},\quad
\mathsf{c}^{(1)}_\e\mathsf{c}^{(3)}_\e\mathsf{c}^{(5)}_\e+
\mathsf{c}^{(2)}_\e\mathsf{c}^{(4)}_\e\mathsf{c}^{(6)}_\e\to\mathsf{c},
\end{equation*}
$\mathsf{c}=\left\|J_{n_p}\left(\sqrt{\l_0}r\right)\phi^+(n_p\th)\right\|^2
=\left\|J_{n_p}\left(\sqrt{\l_0}r\right)\phi^-(n_p\th)\right\|^2$.
The convergences obtained can not hold at the same time, hence,
$\l_\e^{(2p-2)}=\l_\e^{(2p-1)}$. So, if $n_p>0$, then the
associated eigenvalue $\l_\e^{(2p-2)}=\l_\e^{(2p-1)}$ is double.
The perturbed eigenvalue associated with the index $n_p=0$ has
the asymptotics distinct from the asymptotics associated with
other indexes, what means that this eigenvalue is simple.

We proceed to the justification of the asymptotics of the
perturbed eigenfunctions. Let $n_p>0$. We set
\begin{equation*}
\Psi_\e^{(i)}=\sum\limits_{j=2p-2}^{2p-1}\mathsf{b}_\e^{ij}
\psi_\e^{(j)},\quad i=2p-2,2p-1.
\end{equation*}
It is obvious  that $\Psi_\e^{(i)}$ are eigenfunctions of the
perturbed problem associated with the double eigenvalue
$\l_\e^{(2p-2)}$. Due to (\ref{4.14}), the above estimates for
$\widehat{u}_\e^{(i)}$ and the convergence of
$\widetilde{\psi}_{\e,K}^{(i)}$ to $\psi_0^{(i)}$, we obtain
that $\Psi_\e^{(i)}$ converges to $\psi_0^{(i)}$. The assertions
(\ref{4.14}), estimates for $\widehat{u}_\e^{(i)}$ and the
statements of Theorems~\ref{th2.1}~and~\ref{th3.1} for the
functions $R_{\e,K}$ imply the inequalities
\begin{align*}
\left\|\psi_{\e,K}^{(i)}-\Psi_\e^{(i)}\right\|&
\le
\left\|\t\psi_{\e,K}^{(i)}-\Psi_\e^{(i)}\right\|+
\left\|\t\psi_{\e,K}^{(i)}-\psi_\e^{(i)}\right\|\le
\\
{}&\le \left\|\widehat{u}_\e^{(i)}\right\|+
\left\|\t\psi_{\e,K}^{(i)}-\psi_\e^{(i)}\right\|=O\left(\e^{K-4}\right)
\end{align*}
which mean that the asymptotics for the eigenfunctions
associated with the double eigenvalue
$\l_\e^{(2p-2)}=\l_\e^{(2p-1)}$ have the form (\ref{3.1}). For
the simple eigenvalue $\l_\e^{(1)}$ we consider the associated
eigenfunction
\begin{equation*}
\Psi_\e^{(1)}=\mathsf{b}_\e^{11}\psi_\e^{(1)},
\end{equation*}
converging to $J_0\left(\sqrt{\l_0}r\right)$, and by analogy
with the case of double eigenvalue one can easily prove the
estimate
\begin{equation*}
\left\|\psi_{\e,K}^{(1)}-\Psi_\e^{(1)}\right\|=O\left(\e^{K-4}\right),
\end{equation*}
which means that the asymptotics for the eigenfunction
associated with $\l_\e^{(1)}$, has the form (\ref{2.34}). The
proof of Theorem~\ref{th1.1} is complete.

\centsectn{Appendix}

Here we shall prove that for some positive $A$ there exists
$\l_0$ being a root of the equation of the equation (\ref{1.4})
for different $n$ simultaneously. We introduce the notations
$\mathsf{f}_{n,A}(t)=t J'_n(t)+A J_n(t)$, $n\in\mathbb{Z}_+$,
$A\ge 0$, $t\in (0,+\infty)$. The zeroes of the function
$\mathsf{f}_{n,A}(t)$ for nonnegative $A$ are roots of the
equation (\ref{1.4}). Let us prove that there exist $A>0$, $n$,
$m$, $n\not=m$, for those the functions $\mathsf{f}_{n,A}$ and
$\mathsf{f}_{m,A}$ have a common positive root. We set
$\mathsf{F}_{n,m}(t)=t\left(J'_n(t)J_m(t)-J'_m(t)J_n(t)\right)$.
Let some point $t=t_0$ be a root of the function
$\mathsf{F}_{n,m}$ and it is not a zero of the functions $J_n$
and $J_m$. Then it follows from the equality
$t_0\left(J'_n(t_0)J_m(t_0)-J'_m(t_0)J_n(t_0)\right)=0$ that
\begin{equation*}
t_0\frac{J'_n(t_0)}{J_n(t_0)}=t_0\frac{J'_m(t_0)}{J_m(t_0)}.
\end{equation*}
Let $B=t_0{J'_n(t_0)}/{J_n(t_0)}$. If $B\le0$, then the point
$t_0$ is a common root of the functions $\mathsf{f}_{n,A}$ and
$\mathsf{f}_{m,A}$ as $A=-B$. Thus, if we find a root of the
function $\mathsf{F}_{n,m}$ for some  $n$ and $m$ and check the
inequality $t_0{J'_n(t_0)}/{J_n(t_0)}<0$, we shall get the
statement being proved. We make $n=6$, $m=3$. Then
$\mathsf{F}_{6,3}(8)\approx-0.1673037488<0$,
$\mathsf{F}_{6,3}(9)\approx0.0658220035>0$. Since
$\mathsf{F}_{6,3}$ is a smooth function, then there exists a
zero of the function $\mathsf{F}_{6,3}$ in the interval $(8,9)$,
we denote it by $t_0$. Let $j_{p,q}$, $j'_{p,q}$ be positive
roots of $J_p$ and $J'_p$ taken in ascending order:
$j_{p,1}<j_{p,2}<\ldots$, $j'_{p,1}<j'_{p,2}<\ldots$. We have:
$j_{3,1}\approx6.380161896<8$, $j_{3,2}\approx9.761023130>9$,
$j_{6,1}\approx9.936109524>9$, $j'_{6,1}\approx7.501266145<8$.
These equalities imply that there are no zeroes of the functions
$J_3$ and $J_6$ in the interval $(8,9)$. The function $J_6(t)$
is positive for $t\in(0,j_{6,1})$, and $j'_{6,1}<8<9<j_{6,1}$.
For this reason, the inequalities $J_6(t)>0$, $J'_6(t)<0$ are
true as $t\in(8,9)$, from what we deduce that $t
J'_6(t)/J_6(t)<0$ as $t\in(8,9)$. Thus, there exists a zero
$t_0$ of the function $\mathsf{F}_{6,3}$ in the interval
$(8,9)$, that is not a zero of the functions $J_3$ and $J_6$,
moreover, $t_0 J'_6(t_0)/J_6(t_0)<0$. Therefore, $t_0$ is a of
the equation (\ref{1.4}) for $n=6$ and $n=3$ with
$A=-t_0J'_6(t_0)/J_6(t_0)>0$.

\centsectn{Acknowledgments}

The author thanks R.R.~Gadyl'shin for helpful discussion. The
author is supported by RFBR grant, no. 02-01-00693, by the
program ''Leading Scientific Schools'', and partially supported
by Federal Program "Universities of Russia", grant no.
UR.04.01.010

\centsectn{References}

\begin{enumerate}
\def\theenumi{[\arabic{enumi}]}
\item\label{Fr}A.~Friedman, Ch.~Huang and J.~Yong, Effective
permeability of the boundary of a domain, Commun. in Partial
Differential Equations 20 (1995), 59-102.

\item\label{Ch} G.~A.~Chechkin, Averaging of boundary value problems with
singular perturbation of the boundary conditions, Russian Acad.
Sci. Sb. Math. 79 (1994), 191-220.

\item\label{LP} M.~Lobo and E.~Perez, Asymptotic behaviour of an
elastic body with a surface having small stuck regions,
Mathematical Modelling and Numerical Analysis, 22 (1988), 609-624.

\item\label{OC} O.~A.~Oleinik, G.~A.~Chechkin, On boundary-value
problems for elliptic equations with rapidly changing type of
boundary conditions, Russ. Math. Surv. 48 (1993), 173-175.

\item\label{SMJ} R.~R.~Gadyl'shin, G.~A.~Chechkin, A boundary value problem
for the Laplacian with rapidly changing type of boundary
conditions in a multidimensional domain, Sib. Math. J., 40 (1999),
229-244.

\item\label{Dr} G.~A.~Chechkin, E.~I.~Doronina, On asymptotics of
spectrum of boundary value problem with nonperiodic rapidly
alternating boundary conditions, Functional Differential
Equations, 8 (2001), 111-122.

\item\label{DU} R.~R.~Gadyl'shin, The asymptotics of eigenvalues of the
boundary value problem with rapidly oscillating boundary
conditions, Differential equations 35 (1999), 540--551.

\item\label{ME} A.~M.~Il'in, "Matching of asymptotic expansions of
solutions of boundary value problems", Amer. Math. Soc.,
Providence, RI, 1992.

\item\label{CE} M.~I.~Vishik and L.~A.~Lyusternik, Regular degeneration and
the boundary layer for linear differential equations with small
parameter, Amer. Math. Soc. Transl. (2) 35 (1962), 239--364

\item\label{MS} N.~N.~Bogolyubov and Yu.~A.~Mitropol'sk\v i, "Asymptotics
methods in theory of nonlinear oscillations", Gordon and Breach,
New York, 1962.

\item\label{KF} A.~Kratzer und W.~Franz, "Transzendte funcktionen",
Akademische verlagsgesellschaft, Leipzig, 1960.

\item\label{GR}
I.~S.~Gradshtein and I.~M.~Ryzhik, "Tables of Integrals, Sums,
Series, and Products", Fizmatgiz, Moscow, 1989; English transl.,
Acad. Press, New York, 1969.

\end{enumerate}
\end{document}